\documentclass[fleqn,usenatbib]{mnras}

\usepackage{newtxtext,newtxmath}

\usepackage[T1]{fontenc}
\usepackage{ae,aecompl}

\usepackage{graphicx}	
\usepackage{amsmath}	
\usepackage{amssymb}	
\usepackage{multicol}   
\usepackage{longtable} 
\usepackage{caption} 
\usepackage{float} 

\title[Kinematics of the Tucana Dwarf Galaxy]{Kinematics of the Tucana Dwarf Galaxy: An Unusually Dense Dwarf in the Local Group}

\author[A. L. Gregory et al.]{Alexandra L. Gregory$^{1}$\thanks{E-mail: a.gregory@surrey.ac.uk},
Michelle L. M. Collins$^{1}$,
Justin I. Read$^{1}$,
Michael J. Irwin$^{2}$,
\newauthor Rodrigo A. Ibata$^{3}$,
Nicolas F. Martin$^{3, 4}$,
Alan W. McConnachie$^{5}$,
Daniel R. Weisz$^{6}$
\\
$^{1}$Department of Physics, University of Surrey, Guildford, GU2 7XH, Surrey, UK\\
$^{2}$Institute of Astronomy, Madingley Rise, Cambridge CB3 0HA, UK\\
$^{3}$Universit\'e de Strasbourg, CNRS, Observatoire Astronomique de Strasbourg, UMR 7550, F-67000 Strasbourg, France\\
$^{4}$Max-Planck-Institut f\"ur Astronomy, K\"onigstuhl 17, D-69117, Heidelberg, Germany\\
$^{5}$National Research Council, Herzberg Institute of Astrophysics, 5071 West Saanich Road, Victoria, BC V9E 2E7, Canada\\
$^{6}$Department of Astronomy, University of California, Berkeley, CA 94720, USA
}

\date{Accepted XXX. Received YYY; in original form ZZZ}

\pubyear{2019}

\begin{document}
\label{firstpage}
\pagerange{\pageref{firstpage}--\pageref{lastpage}}
\maketitle

\begin{abstract}
\noindent We present new FLAMES$+$GIRAFFE spectroscopy of 36 member stars in the isolated Local Group dwarf spheroidal galaxy Tucana. We measure a systemic velocity for the system of $v_{\mathrm{Tuc}}=216.7_{-2.8}^{+2.9}$kms$^{-1}$, and a velocity dispersion of $\sigma_{\mathrm{v,Tuc}}=14.4_{-2.3}^{+2.8}$kms$^{-1}$. We also detect a rotation gradient of $\frac{dv_{r}}{d\chi}=7.6^{+4.2}_{-4.3}$ kms$^{-1}$ kpc$^{-1}$, which reduces the systemic velocity to $v_{\mathrm{Tuc}}=215.2_{-2.7}^{+2.8}$ kms$^{-1}$ and the velocity dispersion to $\sigma_{v,\mathrm{Tuc}}=13.3_{-2.3}^{+2.7}$ kms$^{-1}$. We perform Jeans modelling of the density profile of Tucana using the line--of--sight velocities of the member stars. We find that it favours a high central density consistent with `pristine' subhalos in $\Lambda$CDM, and a massive dark matter halo ($\sim$$10^{10}$\,M$_\odot$) consistent with expectations from abundance matching. Tucana appears to be significantly more centrally dense than other isolated Local Group dwarfs, making it an ideal laboratory for testing dark matter models.

\end{abstract}




\section{Introduction} \label{sec:intro}
Isolated dwarf galaxies are insulated from tidal effects and interactions, and they therefore provide a unique window onto the role environment plays in galactic evolution. However, whilst there are many detailed studies of Milky Way and Andromeda satellites (e.g. \citealt{simon07,tollerud12,collins13}), comparative observations of isolated dwarfs require a more concerted effort due to the greater distances involved. The ACS LCID project has mapped out the star formation histories of several isolated Local Group dwarfs (e.g. \citealt{monelli10,gallart15,aparicio16}), while the Solo Survey \citep{higgs16} is a wide--field photometric survey targeting isolated dwarfs within 3 Mpc of the Milky Way. Detailed studies of such galaxies are necessary to fully understand galaxy formation and evolution, and may provide answers to several small scale issues with the $\Lambda$ cold dark matter (CDM) paradigm (e.g. \citealt{bullock17}).

Pure dark matter structure formation simulations in $\Lambda$CDM predict that dwarf galaxies should reside in high density halos \citep{dubinski91,navarro96}, with a central dark matter density $\rho_{\rm{DM}}(150 \rm{pc}) > 10^8$ M$_{\odot}$ kpc$^{-3}$ \citep{read18a,read18b}. This has long been known to be inconsistent with observations of gas rich dwarf irregular galaxies (e.g. \citealt{moore94,read17a}), which has become known as the `cusp-core' problem. Similarly, the inferred masses of most satellite dwarf galaxies within their half--light radii are also found to be inconsistent with this prediction \citep{read06b,boylankolchin11,boylankolchin12}. Simulated dark matter subhalos appear to be too dense to host the observed dwarf satellites of the Milky Way and Andromeda (M31), a problem referred to as `Too Big to Fail'.

\begin{table}
	\centering
	\caption{Key Properties of Tucana. a) \citet{lavery92}; b) \citet{bernard09}; c) \citet{saviane96}; d) \citet{mateo98}; e) \citet{hidalgo13}}
	\label{tab:tuc_properties}
	\begin{tabular}{|l|l|}
		\hline
		\textbf{RA$^{a}$} & 22 41 49.6\\[3pt]
		\textbf{Declination$^{a}$} & -64 25 10\\[3pt]
		\textbf{Distance from Milky Way$^{b}$} & 887$\pm$49 kpc\\[3pt]
		\textbf{Core Radius$^{c}$} & 42$\pm$6"\\[3pt]
		& 176$\pm$26 pc\\[3pt]
		\textbf{Half--Light Radius$^{c}$} & 66$\pm$12"\\
		& 284$\pm$54 pc\\[3pt]
		\textbf{Ellipticity$^{c}$} & 0.48$\pm$0.03\\[3pt]
		\textbf{Position Angle of Major Axis$^{c}$} & 97$^{\circ}\pm$2$^{\circ}$\\[3pt]
		\textbf{Luminosity$^{d}$} & 5.5$\times$10$^{5}$ L$_{\odot}$\\[3pt]
		\textbf{Stellar Mass$^{e}$} & 3.2$\times$10$^{6}$ M$_{\odot}$\\ \hline
	\end{tabular}
\end{table}

One elegant solution to both of the above problems is the idea of `dark matter heating' (e.g. \citealt{navarro96,read05,pontzen12,pontzen14,read16}). In this scenario, repeated gas inflow and outflow cause the central gravitational potential of the dwarf galaxy to fluctuate. The dark matter responds to this by migrating outwards, lowering the inner dark matter density. \citet{read18a} have recently found an anti--correlation between the amount of star formation and the central dark matter density in a sample of 16 nearby dwarf galaxies, exactly as expected if dark matter migrates slowly outwards as star formation proceeds. However, their sample of dwarf galaxies is small and may suffer from selection effects. For this reason, it is interesting to measure the inner dark matter density of a larger sample of dwarfs, particularly dwarfs with a purely old stellar population that are expected to retain their `pristine' central dark matter density.

In this paper, we investigate an isolated dwarf spheroidal (dSph) galaxy of the Local Group--- Tucana--- whose star formation shut down long ago. This makes it a particularly clean test for probing the nature of dark matter, since it is less likely to have had its dark matter `heated up' (e.g. \citealt{dicintio14,onorbe15,read16,bermejo18,read18a,read18b}). Tucana was rediscovered and proposed as a Local Group member by \citet{lavery92} after appearances in earlier catalogues. It is located 880 kpc from the Milky Way and 1350 kpc from M31 \citep{castellani96, fraternali09}, making it one of the most isolated galaxies of the Local Group. Observations by \citet{fraternali09} suggest that Tucana is receding from both the Milky Way and the Local Group, and, if bound, has not yet reached apocentre. It is likely to have been isolated for the majority of its history, although tracing the kinematics to higher redshift may imply a possible interaction with the Milky Way around 10 Gyr ago \citep{fraternali09}. \citet{sales07} suggest that Tucana's isolation may be the result of a three body ejection mechanism, potentially involving the Milky Way and the Magellanic Clouds. Along with And XVIII, the other isolated dSph in the Local Group is Cetus, located 775 kpc from the Milky Way \citep{whiting99,lewis07}. Both these isolated dwarfs lie in the direction of Sculptor, and it has been postulated that they may form part of a bridge between the Local Group and the Sculptor Group \citep{whiting99, fraternali09}.

Tucana has experienced no recent star formation. A study by \citet{monelli10} as part of the ACS LCID project found that Tucana experienced a strong burst of star formation $\sim13$ Gyr ago, lasting for ~1 Gyr, with star formation having stopped completely by $\sim9$ Gyr ago, with the exception of a small intermediate age, low metallicity population interpreted as owing to contamination by blue stragglers. The authors show that the colour magnitude diagram of Tucana exhibits the typical features of an old stellar population, such as a lack of a blue main sequence. \citet{gallart15} confirm that $90\%$ of Tucana's stars formed more than 10 Gyr ago. \citet{avilavergara16} find that $75\%$ of Tucana's history has been spent as a `closed box' with no net inflow or outflow of gas. A similar chemical history was also inferred for Cetus \citep{avilavergara16}. It has been proposed that the purported interaction between Tucana and the Milky Way some 10 Gyr ago could have stripped enough gas to completely shut down star formation in the galaxy \citep{teyssier12}. In agreement with the lack of ongoing star formation, \citet{oosterloo96} observe that there is no HI emission within the optical boundary of Tucana. \citet{fraternali09} demonstrate that a nearby (on--sky) HI cloud is more likely to be associated with the Magellanic stream. Through a combination of its unusual location and chemical history, Tucana is a highly unique and interesting Local Group object.

\begin{figure*}
	\centering 
	\includegraphics[width=0.9\linewidth] {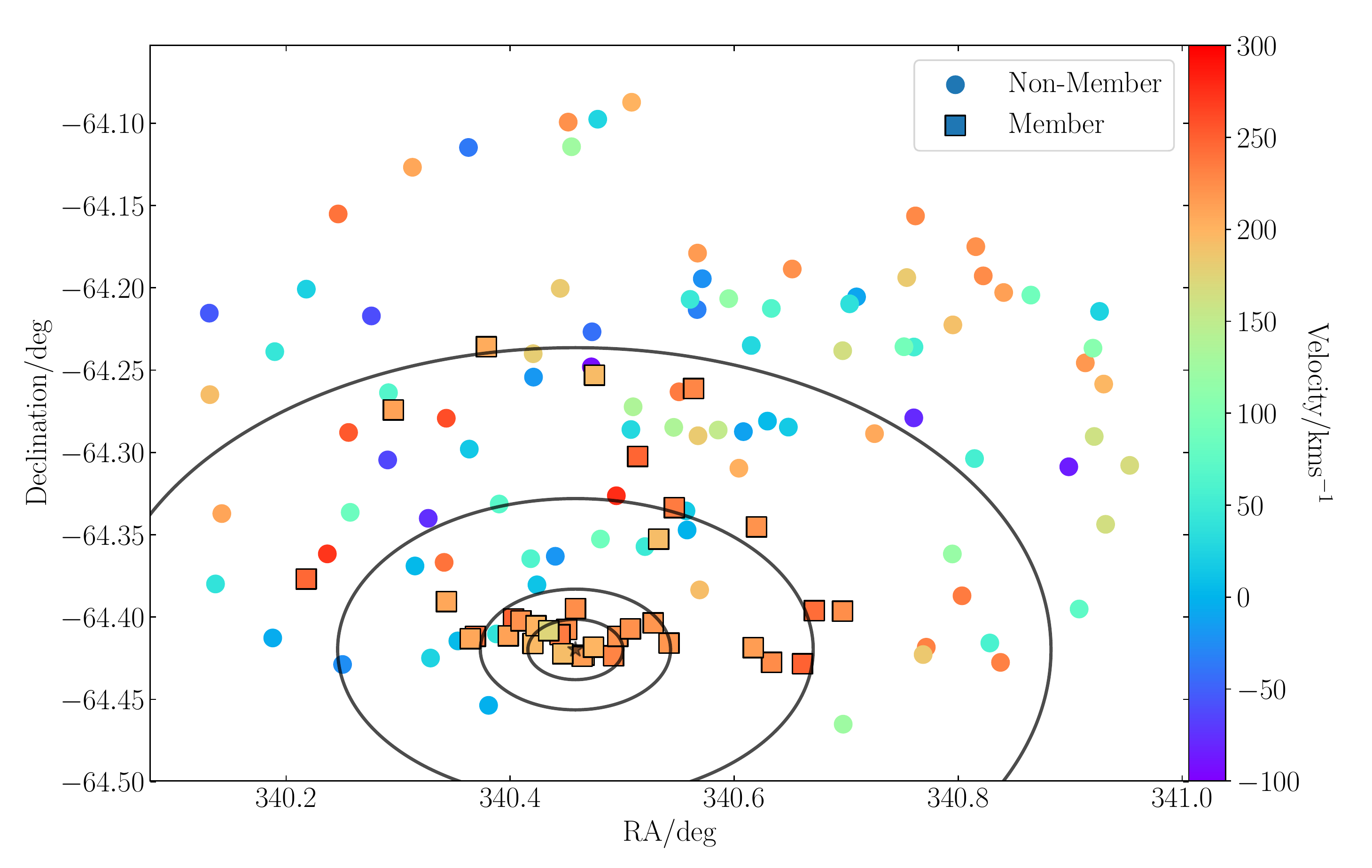}
	\caption{RA--Dec. map of observed objects, plotted relative to the J2000 coordinates of Tucana. Objects are colour coded by their radial velocity. Identified Tucana members are shown as squares. The black circles mark 1$\times$, 2$\times$, 5$\times$ and 10$\times$ the half--light radius of Tucana.}
	\label{fig:radec} 
\end{figure*}

A kinematic study of Tucana was previously undertaken by \citet{fraternali09}, hereafter F09. This study used the FORS2 instrument at the Very Large Telescope (VLT) to measure the radial velocities of red giant branch stars in Tucana, and hence estimate Tucana's systemic velocity and velocity dispersion, constraining the mass and metallicity of the dSph. The stellar radial velocities were found by cross correlating a template with the Ca triplet of each star. 20 stars were identified as candidate members of Tucana based on their velocity. Using a maximum likelihood method to fit a Gaussian profile to the histogram of velocities, the authors obtain a systemic velocity of $v=193.0\pm$4.9 kms$^{-1}$ and a velocity dispersion of $\sigma_{{v}}=17.4^{+4.5}_{-3.5}$ kms$^{-1}$. From this they calculate a mass--to--light ratio of $M/L=105^{+95}_{-49}$, implying a mass of $M\mathrm{_{half}}$$\approx6\times10^{7}$ M$_{\odot}$. F09 also detect a rotation signature of magnitude $v_{\mathrm{rot}}=16$ kms$^{-1}$ along the major axis. Accounting for this slightly increases the systemic velocity to $v=194.0\pm4.3$ kms$^{-1}$, and reduces the velocity dispersion to $\sigma_{v}=15.8^{+4.1}_{-3.1}$ kms$^{-1}$. This corresponds to a system mass of $M\mathrm{_{half}}$$\approx5\times10^{7}$ M$_{\odot}$. F09 also measured a mean metallicity of [Fe/H]=-1.95$\pm$0.15, with a metallicity dispersion of 0.32$\pm$0.06dex.

The aim of our study is to measure the velocity and velocity dispersion of Tucana to a higher accuracy than before. The velocity dispersion measured by F09 is unusually high and may be consistent with the most massive surviving subhalos in pure dark matter $\Lambda$CDM simulations; however the uncertainties on the result are large. We aim to use higher resolution spectroscopy of a larger stellar sample to constrain the velocity dispersion and density profile of Tucana. This paper is structured as follows: Section \ref{sec:obs} describes the observations and data reduction process. We detail the calculation of the velocity dispersion, including determination of the radial velocities and errors, membership probabilities of the Tucana candidates, and calculation of the mass of Tucana in section \ref{sec:veldisp}. Details of the process for modelling the density profile are given in section \ref{sec:gravsphere}. The implications of these results are discussed in section \ref{sec:disc}, and we conclude in section \ref{sec:conclusion}. Key properties of Tucana are listed in Table \ref{tab:tuc_properties}. These values are used throughout this paper unless otherwise stated.

\section{Observations}\label{sec:obs}

Photometric data for Tucana was obtained using the Magellan/ Megacam instrument on the Clay telescope at Las Campanas observatory on 14th November 2012 as part of the Solo (Solitary Local Dwarfs) observing campaign \citep{higgs16}. Magellan/ Megacam is a 9$\times$4 array of pixel CCDs with a pixel scale 0.08 arcsec pix$^{-1}$. Two pointings were targeted, with 3 exposures were stacked in both the $g$--band and $i$--band for each field. Each $g$--band exposure was 150s, for a total integration time of 450s. In the $i$-band the exposures were 300s each, for a total integration time of 900s. Seeing
ranged from 0.68--0.90 arcsec in the $g$--band and 0.55--0.72 in the
$i$--band. The photometry was reduced using the Cambridge Astronomical Survey Unit (CASU) following the process described in \citet{richardson11} and \citet{higgs16}. Including only point--like sources, defined as those with classifiers $c_{g}$, $c_{i}=-1$ or -2, this catalogue featured 7203 objects.

To obtain accurate spectroscopy of the faint stellar population of Tucana, the 8.2m Very Large Telescope (VLT) in Paranal, Chile was used. Observations were taken using the FLAMES$+$GIRAFFE spectrograph over 6 nights through June, August and September 2015. GIRAFFE is a fibre--fed spectrograph for the visible range 3700--9000\AA. The instrument was used in the Medusa mode, allowing observations of up to 132 objects simultaneously, each with an aperture of 1.2" on the sky. Two fibre configurations were used to maximise the number of targets available. Each exposure was 1200s, and 39 exposures were taken in total: 21 in the first configuration for an integration time of 7 hours; and 18 in the second configuration for a 6 hour integration time. The LR8 grating was used, encompassing a 1190\AA\ wavelength band centred on 8817\AA\, which covers the three peaks of the Ca II triplet at around 8500\AA. This setup provides a spectral resolution of $R=6500$, capable of resolving velocity dispersions even in very faint dSphs \citep{koposov11}. Targets were selected using the Magellan photometry. Overall, 165 individual objects were observed, along with 24 sky regions.

The FLAMES$+$GIRAFFE spectrograph is a higher resolution instrument than FORS2 (which was used in F09), with a resolving power of R=6500 as opposed to R=3200. This allows more accurate measurements of stellar radial velocities, potentially reducing the uncertainties by a factor of 2, and thus better constraining the dispersion. F09 observe 23 stars to a S/N suitable for determining velocities, with 20 of these identified as members. We observe a much larger number of stars out to a wider radius than F09, generating a larger catalogue of Tucana members and reducing the uncertainties in the results.

For the reduction, we used the pipeline for the GIRAFFE instrument provided by ESO. The raw science data was reduced using the \texttt{giscience} recipe via the graphical interface \texttt{Gasgano}. The pipeline provides corrections for detector effects, including dark and bias corrections, then traces the fibre positions and aligns each fibre with the corresponding spectrum. The spectra are output along with a wavelength calibration and descriptors of the fibre setup and observation. Where possible, calibration frames taken alongside the observations were used (standard calibration files were used for PSF$\_$WIDTH and PSF$\_$CENTROID, which provide the width and centre of the fitted fibre profile). 

All spectra were normalised to have unity continuum flux by dividing through by the continuum level. For each observation, the individual sky spectra were median combined to generate a master sky spectrum for each pointing. A median average was chosen as it removes any spurious lines present in individual sky spectra from the master. In order to obtain an accurate sky subtraction, both science and sky were shifted to a continuum level of 0, and the master sky for the relevant pointing normalised by scaling the height of the sky line at 8500\AA\ to match the height of the same sky line in the science spectrum. The sky template was then subtracted from the science spectrum. The resulting spectrum was shifted back to a continuum level of one, and a heliocentric correction was applied to the wavelength scale. Finally, all the spectra of a given object were median combined to give one master spectrum for each object observed. The overall data reduction process results in a set of 165 sky subtracted, heliocentric corrected stellar spectra from Tucana.

\section{Determination of the Velocity Dispersion}\label{sec:veldisp}

\begin{figure}
	\centering
	\includegraphics[width=\linewidth] {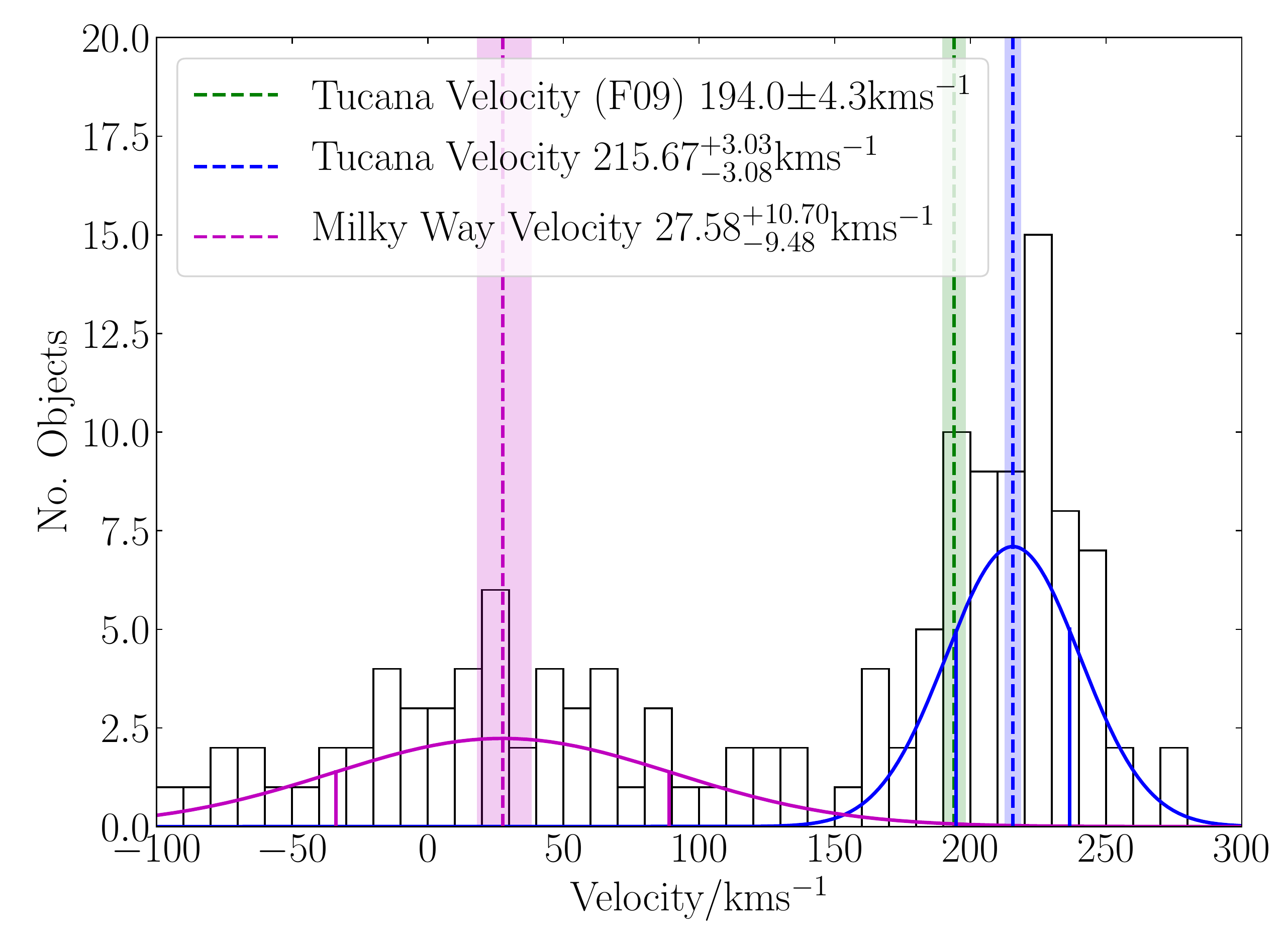}
	\caption{Histogram of the radial velocities of all objects, except those with a cross correlation failure. The systemic velocity of Tucana as determined by F09 is shown by the green dashed line, with its uncertainty range shaded, as are the fitted velocities of the two peaks representing Tucana and the Milky Way.}
	\label{fig:histogram}
\end{figure}

\subsection{Radial Velocities and Errors}\label{sec:radvel}

The line--of--sight velocities of the stars in Tucana were found by cross correlating the spectrum for each object with a template spectrum. The template used was a Gaussian model of the rest Ca II triplet, with peaks at 8499\AA, 8543\AA\ and 8663\AA, retaining the relative equivalent widths of the lines. The cross correlation function was calculated using the \texttt{pyasl.crosscorrRV} function from PyAstronomy\footnote{https://github.com/sczesla/PyAstronomy}. Many spectra exhibit obscuring noise around the third Ca II triplet line; to ensure an accurate result, we therefore only use the first two lines in the cross correlation. Based on the expected velocity of Tucana (following the results of F09) and the Milky Way (from a Besan\c{c}on model--- see section \ref{sec:pvel}), the allowed velocities were limited to the range -100 kms$^{-1}<v<$300 kms$^{-1}$. The velocity where the cross correlation function is maximised is then taken to be the radial velocity of the object.

Errors on the radial velocities were found by following the iterative Monte Carlo process outlined in \citet{tollerud12}. 1000 iterations of each object spectrum were generated by adding noise seeded by the variance per pixel, assuming independent, Poisson distributed noise. Each iteration is cross correlated with the template Ca II triplet, and all 1000 velocities are plotted as a histogram. A Gaussian profile is fitted to the primary peak of the histogram, and the best fit mean and standard deviation are taken as the radial velocity and corresponding error for the given object. There are 25 objects for which the low S/N of the spectrum leads to a cross correlation failure, and 9 for which a Gaussian could not be fitted to the profile. These are removed from the dataset to leave 131 successfully reduced objects. The average S/N of these spectra is 8.4/pixel (with a pixel size of 0.2\AA\ /pixel).

\begin{figure*}
	\centering
	\includegraphics[width=0.9\linewidth] {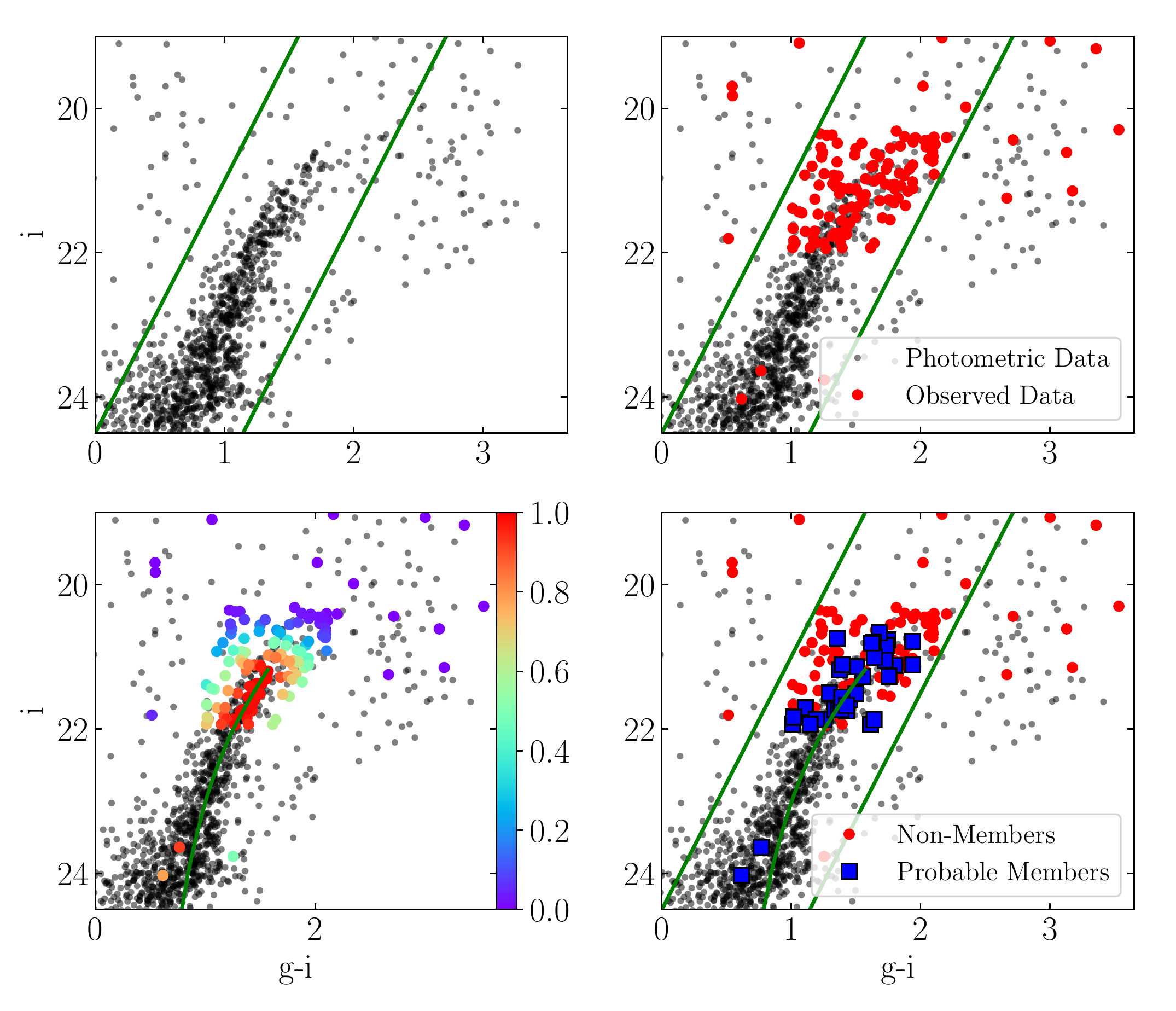}
	\caption{Upper: Colour magnitude diagram for the photometric sources within 4 arcmin of Tucana (left), and with the observed objects overlaid (right). Lower: Colour magnitude diagram overlaid with the 11 Gyr, [$\alpha$/Fe]=0, [Fe/H]=-1.5dex isochrone. Left, the sources from the master catalogue are colour coded by probability of membership. Right, sources identified as members of Tucana (based on both CMD position and velocity) are represented by blue squares, and likely non members by red points. The green lines mark a fiducial `bounding box' indicating the location of the RGB.}
	\label{fig:CMD}
\end{figure*}

Details of the observed objects are listed in table \ref{tab:allobj}, along with the details of the corresponding F09 object where applicable. The positions of these objects relative to Tucana are shown in Fig. \ref{fig:radec}, colour coded by radial velocity. The velocity map indicates a concentration of stars with velocities near the systemic velocity of Tucana in the dense centre of the galaxy. The velocities are plotted as a histogram in Fig. \ref{fig:histogram}, exhibiting well defined Tucana and Milky Way peaks (see section \ref{sec:pvel}).

\subsection{Determination of Membership}\label{sec:memprob}

To establish the true velocity profile of Tucana, we must robustly determine which objects are members of Tucana. To do so, we use elements of the probabilistic method outlined in \citet{collins13}; namely, position on the colour--magnitude diagram, distance from the centre of Tucana, and the velocity of the object. The probability of membership for a given object is defined by

\begin{eqnarray}
P_{i}\propto P_{\mathrm{CMD}} \times P_{\mathrm{dist}} \times P_{\mathrm{vel}}.
\label{eq:probability}
\end{eqnarray}

\noindent Below we outline the method for determining $P\mathrm{_{CMD}}$, $P\mathrm{_{dist}}$ and $P\mathrm{_{vel}}$, and hence $P_{i}$.

\subsubsection{Membership Probability Based on CMD Position} \label{sec:pcmd}

The stellar members of Tucana present in our dataset are expected to lie on the red giant branch (RGB) of the colour--magnitude diagram (CMD). By following the method of \citet{tollerud12} and overlaying appropriate isochrones onto the colour--magnitude diagram, the proximity of each object to the isochrone can be measured using equation \ref{eq:isomem},

\begin{eqnarray}
P\mathrm{_{CMD}}=\exp\bigg[-\bigg(\frac{\Delta(g-i)^2}{2\sigma_c}-\frac{\Delta(i)^2}{2\sigma_m}\bigg)\bigg],
\label{eq:isomem}
\end{eqnarray}

\noindent where $g-i$ is the difference in $g$--band and $i$--band magnitudes, $i$ is the $i$--band magnitude, and $\sigma_c$ and $\sigma_m$ are free parameters which take account of distance and photometry factors. \citet{tollerud12}'s value of $\sigma_c=0.1$ was used, but we used $\sigma_m=0.1$ instead of $\sigma_m=0.5$, as this gave a better fit to the Tucana CMD. $P\mathrm{_{CMD}}$ serves as a proxy for the probability of membership.


\begin{table*}
	\centering
	\caption{Details of all successfully reduced targets observed with FLAMES+GIRAFFE. Where applicable, the corresponding object details from \citet{fraternali09} are also provided. Columns are: (1) Object ID; (2) Line--of--sight heliocentric velocity with error; (3) Right Ascension in J2000; (4) Declination in J2000; (5) \textit{g}--band magnitude from Magellan/ MegaCam imaging; (6) \textit{i}--band magnitude from Magellan/ MegaCam imaging; (7) S/N ratio in pix$^{-1}$; (8) Member?; (9) ID of counterpart in F09 dataset; (10) Velocity and error of F09 counterpart.}
	\label{tab:allobj}
	\begin{tabular}{|c|c|c|c|c|c|c|c|c|c|}
		\hline
		\textbf{Object} & \textbf{Velocity (kms$^{-1}$)} & \textbf{RA (deg)}   & \textbf{Dec. (deg)} & \textbf{\textit{g}} & \textbf{\textit{i}} & \textbf{S/N (pix$^{-1}$)} & \textbf{M?} & \textbf{F09} & \textbf{F09 Vel. (kms$^{-1}$)} \\ \hline
		21138  & 116.7$\pm$30.0 & 340.5952 & -64.2066 & 23.05 & 21.15 & 13.8 & N & -- & -- \\
		51120  & 240.7$\pm$8.4  & 340.3689 & -64.4116 & 22.82 & 21.71 & 10.4 & Y & -- & -- \\
		34818  & -63.0$\pm$3.6  & 340.2907 & -64.3046 & 22.61 & 21.11 & 10.6 & N & -- & -- \\
		...  & ...  & ... & ... & ... & ... & ... & ... & ... & ... \\
       	100021 & 192.7$\pm$13.3 & 340.5692 & -64.3835 & 22.50 & 20.40 & 10.8 & N & 22 & 201.6$\pm$11.5 \\
        ...  & ...  & ... & ... & ... & ... & ... & ... & ... & ... \\
\hline
	\end{tabular}
\flushleft (This table is available in its entirety in a machine-readable form in the online journal. A portion is shown here for guidance regarding its form and content.)
\end{table*}

A CMD was generated for Tucana as shown in the first panel of Fig. \ref{fig:CMD}, using the Magellan photometry. To reduce the density of sources around the RGB, only those sources within 4 arcmin of the centre of Tucana were used. A fiducial bounding box is marked to highlight the position of the RGB. The objects in our dataset were matched against the photometric data using on--sky position in order to obtain the photometric properties of each object. Given that the photometric data was used for targeting the spectroscopic observations, most objects had a near direct match in RA/ Dec. These data are overlaid onto Tucana's CMD in the upper right hand plot of \ref{fig:CMD}. As expected, the majority of catalogue sources appear to lie directly on the RGB.

Isochrones were taken from the Dartmouth Stellar Evolution database \citep{dotter08}, and overlaid onto this CMD. Visual inspection suggests that the isochrone which best represents the RGB of Tucana is an isochrone of age 11 Gyr, [$\alpha$/Fe]=0, and [Fe/H]=-1.5 dex, which is also in agreement with previous measurements \citep{saviane96, fraternali09}. Note that the isochrone is purely used as a guide for selecting the most probable members, and so a formal fitting procedure is not required. The isochrone is shifted by the distance modulus $m-M=24.7$, to account for Tucana's distance of $D=887$ kpc. Equation \ref{eq:isomem} is used to determine the proximity of each object to this isochrone. The lower left hand plot of Fig. \ref{fig:CMD} shows the CMD of photometric sources with the 11 Gyr, [$\alpha$/Fe]=0, [Fe/H]=-1.5dex isochrone overlaid. The observed objects are plotted, colour--coded by probability based on isochrone proximity.

\subsubsection{Membership Probability Based on Velocity} \label{sec:pvel}

We establish the probability of membership based on the velocity of a given object using a maximum likelihood method, first laid out in \citet{martin07} and adapted from \citet{collins13}. The radial velocities of all objects  are plotted as a histogram in Fig. \ref{fig:histogram}. This shows a peak at $v$$\approx$0 kms$^{-1}$ comprising stars in the Milky Way, and a second peak at $v$$\approx$200 kms$^{-1}$ representing the Tucana population. Each peak is well represented by a Gaussian function of the form

\begin{multline}
    P\mathrm{_{peak}}=\frac{1}{\sqrt{2\pi}\sqrt{\sigma_{v,\mathrm{peak}}^2+v_{\mathrm{err}, i}^2}} \\ 
    \times\exp\bigg(-\frac{1}{2}\bigg[\frac{v\mathrm{_{peak}}-v_i}{\sqrt{\sigma_{v,\mathrm{peak}}^2+v_{\mathrm{err}, i}^2}}\bigg]^2\bigg),
\label{eq:probmw}
\end{multline}

\noindent where $v_i$ is the velocity of the given star, $v_{\mathrm{err}}$ is the uncertainty on its velocity, $v_{\mathrm{peak}}$ is the prior on the velocity of the peak, and $\sigma_{v,\mathrm{peak}}$ is the prior on the velocity dispersion. If $P_{\mathrm{Tuc}}$ is the probability of membership of the Tucana peak, and $P_{\mathrm{MW}}$ is the probability of membership of the Milky Way peak, the overall likelihood function becomes

\begin{eqnarray}
    \mathcal L=\sum_{i=1}^{N} \log(\eta\mathrm{_{MW}} P_{\mathrm{MW},i} + \eta\mathrm{_{Tuc}} P_{\mathrm{Tuc},i}),
\label{eq:likelihood}
\end{eqnarray}

\noindent where $\eta_{\mathrm{peak}}$ is the fraction of the stars which belong to that peak. Using Bayesian techniques, the probability that a star belongs to Tucana based on its velocity is then given by

\begin{eqnarray}
P\mathrm{_{vel}}=\frac{P_{\mathrm{Tuc},i}}{P_{\mathrm{MW},i} + P_{\mathrm{Tuc},i}}.
\label{eq:pvel}
\end{eqnarray}

\noindent To measure the velocity and dispersion of Tucana, an MCMC routine is used to fit equation \ref{eq:probmw} to the Tucana peak. For this the \texttt{emcee} python package from \citet{foremanmackey13} is used. The values of the systemic velocity and dispersion of Tucana published in F09 are used as initial values for $v_{\mathrm{Tuc}}$ and $\sigma_{v,\mathrm{Tuc}}$. The velocity of the Milky Way background in this region of the sky is determined using the Besan\c{c}on model \citep{robin03}, which established that Milky Way contaminants are expected to have velocities in the range -50 kms$^{-1}<v<$50 kms$^{-1}$. Therefore, $v_{\mathrm{MW}}=0$ kms$^{-1}$ is taken as the initial value for the velocity of this peak, and $\sigma_{v,\mathrm{MW}}=50$ kms$^{-1}$ as the initial value of the velocity dispersion. The analysis uses 100 walkers taking 30000 steps, with a burn in of 6000 steps. Stars with a radial velocity $v>$295 kms$^{-1}$ or $v<$-95 kms$^{-1}$ are rejected, as these velocities lie at the limits of the accepted range for the cross correlation, and so are likely to indicate a cross correlation failure. The initial values of $\eta_{\mathrm{MW}}$ and $\eta_{\mathrm{Tuc}}$ are both taken to be 0.5, based on a visual inspection of the histogram shown in Fig. \ref{fig:histogram}, and are normalised to sum to 1. The process generates posterior values for the velocity, velocity dispersion and fraction of stars in each peak. 

This establishes a velocity profile for the two peaks, with the velocity, velocity dispersion and fraction of member stars defined. Including all observed objects, the Tucana peak is found to have a velocity of $v_{\mathrm{Tuc}}=215.7_{-3.1}^{+3.0}$ kms$^{-1}$, and a dispersion of $\sigma_{v,\mathrm{Tuc}}=20.9_{-2.5}^{+2.9}$ kms$^{-1}$, while the Milky Way peak has a velocity of $v_{\mathrm{MW}}=27.5_{-9.5}^{+10.7}$ kms$^{-1}$, and a dispersion of $\sigma_{v,\mathrm{MW}}=61.4_{-7.4}^{+9.5}$ kms$^{-1}$. The quoted uncertainties on these values are the 1 sigma uncertainties returned by the MCMC routine These results are used to produce the Gaussian fits plotted in Fig. \ref{fig:histogram}. This velocity is offset from the value measured by F09, and the velocity dispersion is significantly higher. However, this velocity dispersion is artificially increased by the inclusion of non--member stars in the Tucana peak.

We insert the results of the MCMC routine, along with the individual object velocities, into equation \ref{eq:probmw} to calculate the probability of each object belonging to the Tucana peak and the Milky Way peak respectively. These probabilities are then inserted into equation \ref{eq:pvel} to determine the probability of membership of the given object based on velocity.

\subsubsection{Membership Probability Based on Distance} \label{sec:pdist}

\begin{figure}
	\centering
	\includegraphics [width=\linewidth] {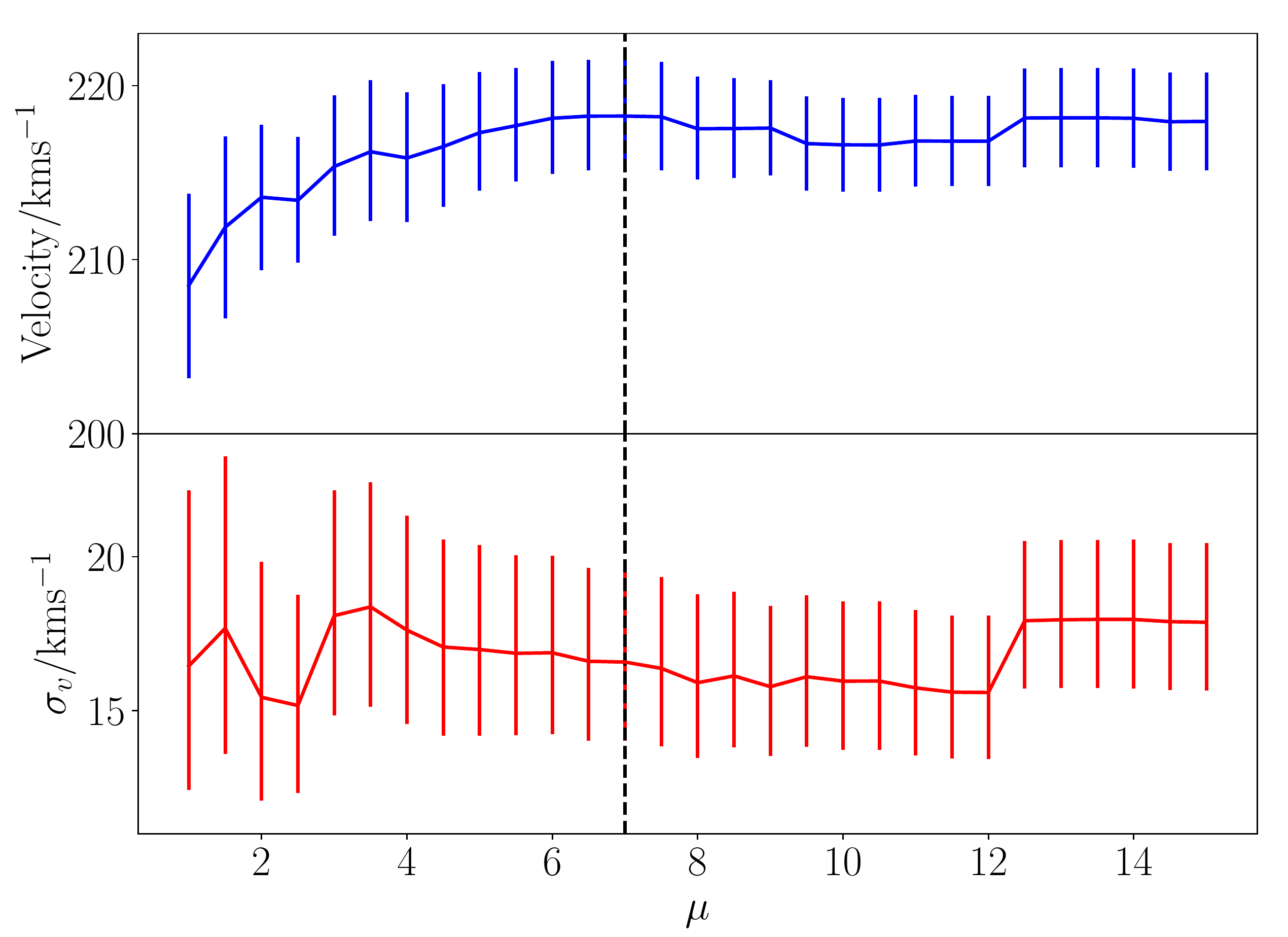}
	\caption{Plot showing how the measured systemic velocity and velocity dispersion of Tucana vary as a function of $\mu$, a multiplicative weight applied to the half--light radius in P$_{\rm{dist}}$. The dashed line marks the optimum value used in our analysis. The systemic velocity and velocity dispersion are generally stable, even at large $\mu$.}
	\label{fig:eta_test}
\end{figure}

We also introduce a parameter to account for the distance of the object to the centre of Tucana. This factor is introduced to ensure that objects with a high probability of membership based on their velocity and CMD position, but which are located far outside the centre of the galaxy, are not included as definite members. From \citet{tollerud12}, the probability that a star is a Tucana member based on its distance from the centre of the galaxy is given by

\begin{eqnarray}
P\mathrm{_{dist}}=\exp\bigg[-\bigg(\frac{\Delta\alpha^{2}+\Delta\delta^{2}}{2\eta r\mathrm{_{half}}^{2}}\bigg)\bigg],
\label{eq:probdist}
\end{eqnarray}

\noindent where $r\mathrm{_{half}}$ is the half--light radius of Tucana, $\Delta\alpha^{2}$ and $\Delta\delta^{2}$ are the distances from the object to the centre of Tucana in RA and declination respectively, and $\eta=1.5$ is a free parameter. We use the literature value of $r\mathrm{_{half}}$.

This expression assumes that all member stars will lie within the half--light radius, and hence is designed to give extra weighting to objects within $r\mathrm{_{half}}$. In our dataset, most objects are located outside the half--light radius, as shown in Fig. \ref{fig:radec}, such that equation \ref{eq:probdist} assigns a very low probability of membership to the majority of our dataset. Therefore, to increase the weighting of stars outside the half--light radius, we introduce a multiplicative factor $\mu$ to our definition of $r\mathrm{_{half}}$, such that equation \ref{eq:probdist} becomes

\begin{eqnarray}
P\mathrm{_{dist}}=\exp\bigg[-\bigg(\frac{\Delta\alpha^{2}+\Delta\delta^{2}}{2\eta (\mu r\mathrm{_{half}})^{2}}\bigg)\bigg].
\label{eq:probdist2}
\end{eqnarray}

\noindent In doing so, we are effectively assuming that most observed members lie within $\mu r\mathrm{_{half}}$ of the centre of Tucana, thus ensuring a more gradual decline in membership probability with distance. To find the optimum value of $\mu$, we allow it to vary between 1 and 15, and measure the systemic velocity and velocity dispersion of the resulting sample, as shown in Fig. \ref{fig:eta_test}. We find that the systemic velocity and velocity dispersion are reasonably stable, even at large $\mu$, because the kinematics of Tucana are well separated from the Milky Way contaminants. The optimum value is $\mu=7.0$, which we use in equation \ref{eq:probdist2} to obtain $P\mathrm{_{dist}}$. Note that this equation is only used to define our membership sample; it is not influenced by and has no effect on the surface brightness profile used in section \ref{sec:gravsphere}.

\bigskip
\noindent The probability of membership based on velocity, $P_{\mathrm{vel}}$, is combined with the probabilities based on CMD position, $P\mathrm{_{CMD}}$, and distance, $P\mathrm{_{dist}}$, using equation \ref{eq:probability}, to determine the probability of membership of each object. By removing objects with a membership probability $P_{i}<0.15$, we obtain a population of 37 member stars, details of which are provided in Table \ref{tab:member_stars}. We compare this sample to the dataset generated by F09. Our data was matched to the objects listed in Table 3 of F09 using on--sky position. Table 3 consists of 23 observed objects, 20 of which are identified as members of Tucana. Position matching found that 13 of these members were present in both datasets. Given the significant difference between the systemic velocities obtained by each study, we do not combine the datasets (see section \ref{sec:f09comparison} for full discussion).

\begin{figure}
	\centering
	\includegraphics[width=\linewidth] {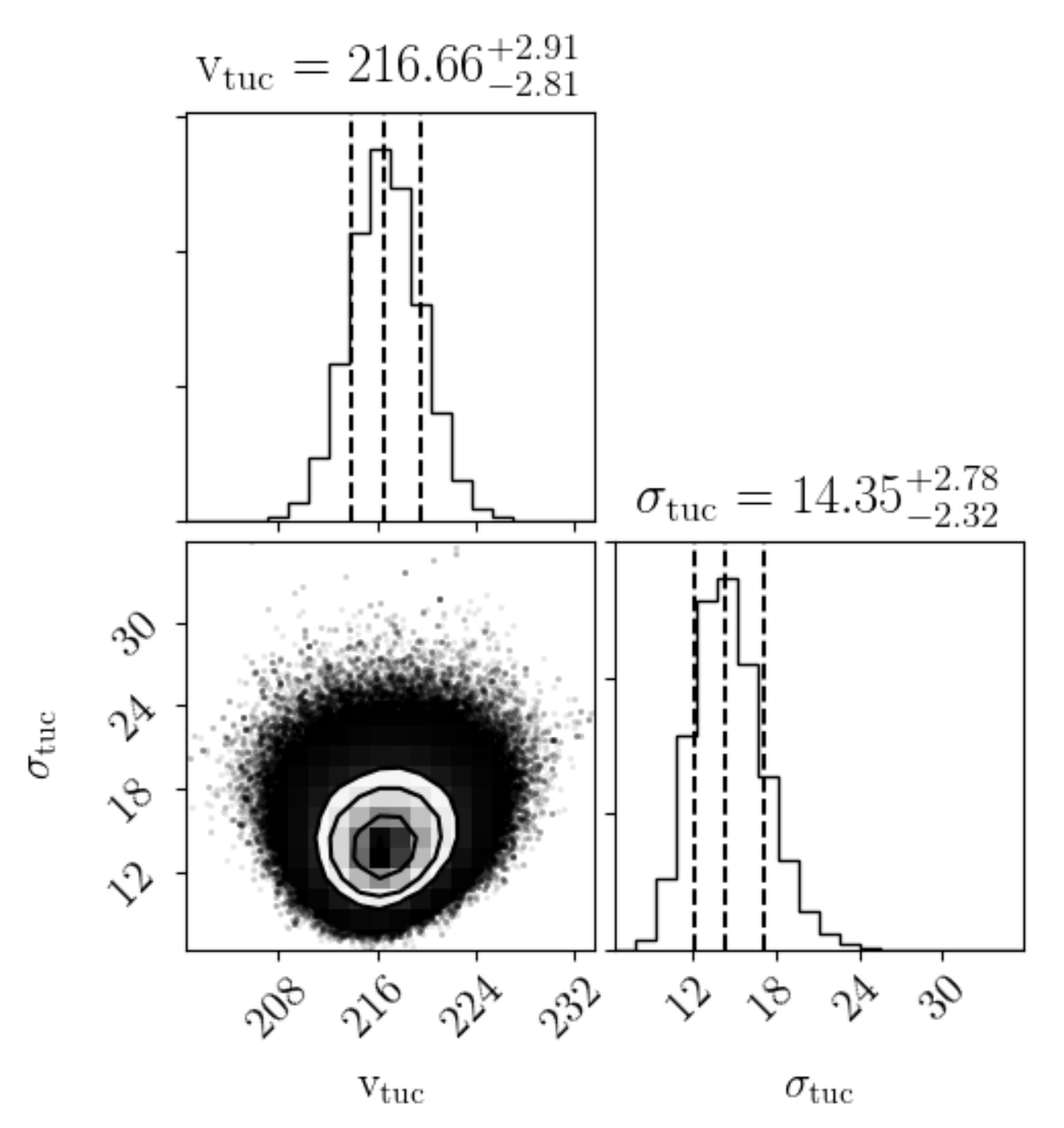}
	\caption{Two--dimensional and marginalised PDFs for the systemic velocity and velocity dispersion (both in kms$^{-1}$) of the identified Tucana members, assuming a purely dispersion supported system. The dashed lines represent the mean value and 1$\sigma$ uncertainties.}
	\label{fig:cornerplot2}
\end{figure}

\begin{table*}
	\centering
	\caption{Details of the 36 identified members of Tucana observed with FLAMES+GIRAFFE. Details of F09 counterparts are listed in table \ref{tab:allobj} where applicable. Columns are: (1) Object ID; (2) Line--of--sight heliocentric velocity with velocity error; (3) Right Ascension in J2000; (4) Declination in J2000; (5) On--sky separation of the object from the central coordinates of Tucana as listed in Table \ref{tab:tuc_properties}; (6) \textit{g}-band magnitude from Magellan/ MegaCam imaging; (7) \textit{i}--band magnitude from Magellan/ MegaCam imaging; (8) S/N ratio in pix$^{-1}$.}
	\label{tab:member_stars}
	\begin{tabular}{|c|c|c|c|c|c|c|c|}
		\hline
		\textbf{Object ID} & \textbf{Velocity (kms$^{-1}$)} & \textbf{RA (deg)}   & \textbf{Declination (deg)} & \textbf{Radius (arcsec)} & \textbf{\textit{g}} & \textbf{\textit{i}} & \textbf{S/N (pix$^{-1}$)}\\ \hline
		51120  & 240.7$\pm$8.4  & 340.3689 & -64.4116 & 141.96 & 22.82 & 21.71 & 10.4 \\
		48384  & 242.4$\pm$8.2  & 340.6717 & -64.3961 & 342.56 & 22.92 & 21.12 & 13.3 \\
		49196  & 249.1$\pm$12.3 & 340.4029 & -64.4011 & 109.11 & 23.12 & 21.86 & 13.4 \\
		100005 & 202.3$\pm$5.5  & 340.4345 & -64.4083 & 55.21  & 22.43 & 20.81 & 8.9  \\
		27260  & 194.1$\pm$12.7 & 340.4753 & -64.2531 & 600.41 & 22.83 & 21.27 & 9.4  \\
		100017 & 211.5$\pm$3.6  & 340.3983 & -64.4114 & 97.94  & 22.52 & 20.77 & 10.4 \\
		100006 & 231.2$\pm$12.3 & 340.4922 & -64.4236 & 54.61  & 22.55 & 21.18 & 12.7 \\
		100009 & 221.8$\pm$4.2  & 340.4961 & -64.4117 & 65.44  & 22.58 & 20.84 & 12.5 \\
		100007 & 196.7$\pm$4.5  & 340.4203 & -64.4161 & 60.52  & 22.77 & 21.05 & 7.8  \\
		50350  & 224.5$\pm$4.2  & 340.4505 & -64.4075 & 45.56  & 22.80 & 21.50 & 9.6  \\
		40705  & 220.6$\pm$10.0 & 340.6201 & -64.3452 & 367.98 & 23.55 & 21.93 & 12.3 \\
		54894  & 248.7$\pm$15.3 & 340.6610 & -64.4284 & 316.65 & 23.04 & 21.59 & 10.9 \\
		50951  & 234.8$\pm$4.3  & 340.4445 & -64.4107 & 38.76  & 22.94 & 21.93 & 11.4 \\
		53700  & 198.5$\pm$5.1  & 340.4661 & -64.4231 & 17.28  & 23.07 & 21.73 & 9.0  \\
		51422  & 208.6$\pm$2.6  & 340.3645 & -64.4132 & 147.77 & 23.17 & 21.75 & 8.8  \\
		100013 & 222.4$\pm$20.8 & 340.5074 & -64.4072 & 88.65  & 22.34 & 20.66 & 12.9 \\
		100002 & 215.2$\pm$5.1  & 340.4645 & -64.4238 & 17.61  & 24.64 & 24.03 & 9.9  \\
		52007  & 219.9$\pm$3.3  & 340.5419 & -64.4159 & 130.64 & 23.06 & 21.86 & 9.5  \\
		30380  & 212.4$\pm$10.8 & 340.2958 & -64.2742 & 581.93 & 23.01 & 21.51 & 13.1 \\
		100016 & 223.0$\pm$6.7  & 340.4097 & -64.4023 & 98.24  & 22.43 & 20.79 & 10.1 \\
		48450  & 224.7$\pm$9.9  & 340.6969 & -64.3965 & 380.28 & 22.85 & 21.83 & 12.1 \\
		100001 & 193.3$\pm$13.6 & 340.4470 & -64.4223 & 19.97  & 22.72 & 20.78 & 12.5 \\
		39027  & 235.7$\pm$18.1 & 340.5467 & -64.3336 & 339.29 & 23.05 & 21.11 & 12.2 \\
		100003 & 199.1$\pm$8.9  & 340.4745 & -64.4183 & 25.74  & 24.40 & 23.64 & 8.1  \\
		45504  & 246.5$\pm$21.2 & 340.2180 & -64.3769 & 404.36 & 23.02 & 21.27 & 12.0 \\
		49630  & 220.1$\pm$5.7  & 340.5278 & -64.4035 & 122.74 & 23.09 & 21.73 & 10.4 \\
		24956  & 205.4$\pm$19.5 & 340.3787 & -64.2358 & 673.64 & 23.51 & 21.87 & 13.1 \\
		100014 & 224.6$\pm$10.1 & 340.4582 & -64.3949 & 89.42  & 22.10 & 20.74 & 13.0 \\
		28435  & 228.4$\pm$9.2  & 340.5639 & -64.2612 & 593.80 & 23.11 & 21.73 & 9.2  \\
		41813  & 193.1$\pm$2.9  & 340.5327 & -64.3526 & 267.91 & 22.64 & 21.14 & 14.0 \\
		47569  & 209.1$\pm$3.4  & 340.3431 & -64.3906 & 207.54 & 22.50 & 21.11 & 13.8 \\
		54691  & 226.4$\pm$11.6 & 340.6335 & -64.4276 & 273.76 & 23.08 & 21.93 & 10.6 \\
		49958  & 201.0$\pm$0.4  & 340.4230 & -64.4053 & 75.50  & 22.95 & 21.56 & 10.7 \\
		34523  & 247.6$\pm$16.3 & 340.5140 & -64.3024 & 431.24 & 22.65 & 21.01 & 9.9  \\
		52613  & 214.1$\pm$7.2  & 340.6170 & -64.4186 & 246.79 & 23.10 & 21.67 & 9.5  \\
		50487  & 174.5$\pm$11.6 & 340.4345 & -64.4083 & 55.21  & 22.43 & 20.81 & 9.0 \\ \hline
	\end{tabular} 
\end{table*}

\subsection{Systemic Velocity and Velocity Dispersion} \label{sysvel}
To generate the final Gaussian fit, we applied an adapted MCMC routine to the 37 identified member stars. We adjust the likelihood function to become

\begin{eqnarray}
\mathcal L=\sum_{i=1}^{N} P_{i} \eta\mathrm{_{Tuc}} P_{\mathrm{Tuc},i},
\label{eq:likelihood2}
\end{eqnarray}

\noindent thus accounting for the probability of membership in the final routine. This returns a final systemic velocity of Tucana of $v_{\mathrm{Tuc}}=218.3_{-3.1}^{+3.2}$ kms$^{-1}$, and a velocity dispersion of $\sigma_{v,\mathrm{Tuc}}=16.6_{-2.6}^{+3.1}$ kms$^{-1}$. The velocity dispersion is now found to be within 1$\sigma$ of the value calculated by F09, but with significantly smaller error bars. The decrease in the size of the uncertainties is consistent with the increase in sample size relative to F09.

Our probabilistic method returns member stars out to $\sim10r_{\rm{half}}$. This is noted as being a particularly large radius at which to find spectroscopically confirmed members. However, it is not without precedent, as \citet{walker09b} find spectroscopic members of Leo V beyond $10r_{\rm{half}}$. In addition, we have shown that the systemic velocity and velocity dispersion are stable to our choice of calibration of the distance probability (see Fig. \ref{fig:eta_test}). Given their strong probabilities of membership, we therefore choose to retain the more distant stars in our membership sample.

One identified member, object ID 37852, has a particularly high velocity of $v_i=277.7\pm12.8$ kms$^{-1}$, some 60 kms$^{-1}$ higher than the measured systemic velocity of Tucana, and hence could be considered a potential contaminant in the sample. It has no counterpart in the F09 dataset, so we cannot make a direct comparison of the velocity measurements, although it does not appear to have significant error bars compared to the rest of our sample. The object is located 5.68 arcmin from the centre of Tucana. Given that it lies $>3\sigma_{v}$ from the mean velocity, we choose to remove this object from our membership sample, in order to avoid including any spurious members. Removing object 37852 reduces the velocity dispersion to $\sigma_{v,\mathrm{Tuc}}=14.4^{+2.8}_{-2.3}$ kms$^{-1}$, and the systemic velocity is reduced to $v_{\mathrm{Tuc}}=216.7^{+2.9}_{-2.8}$ kms$^{-1}$. The  1-- and 2-- dimensional probability distributions for the velocity and velocity dispersion are shown in Fig. \ref{fig:cornerplot2}. These values are smaller than the results from the full sample, but fall within the uncertainties. We therefore use this sample of 36 member stars for the remainder of the analysis. We note that there is still a strong probability that object 37852 is a member of Tucana, but choose to be conservative in our membership definition so as not to overstate our results. Details of object 37852 are included in table \ref{tab:allobj}.

\begin{figure}
	\centering
	\includegraphics[width=\linewidth] {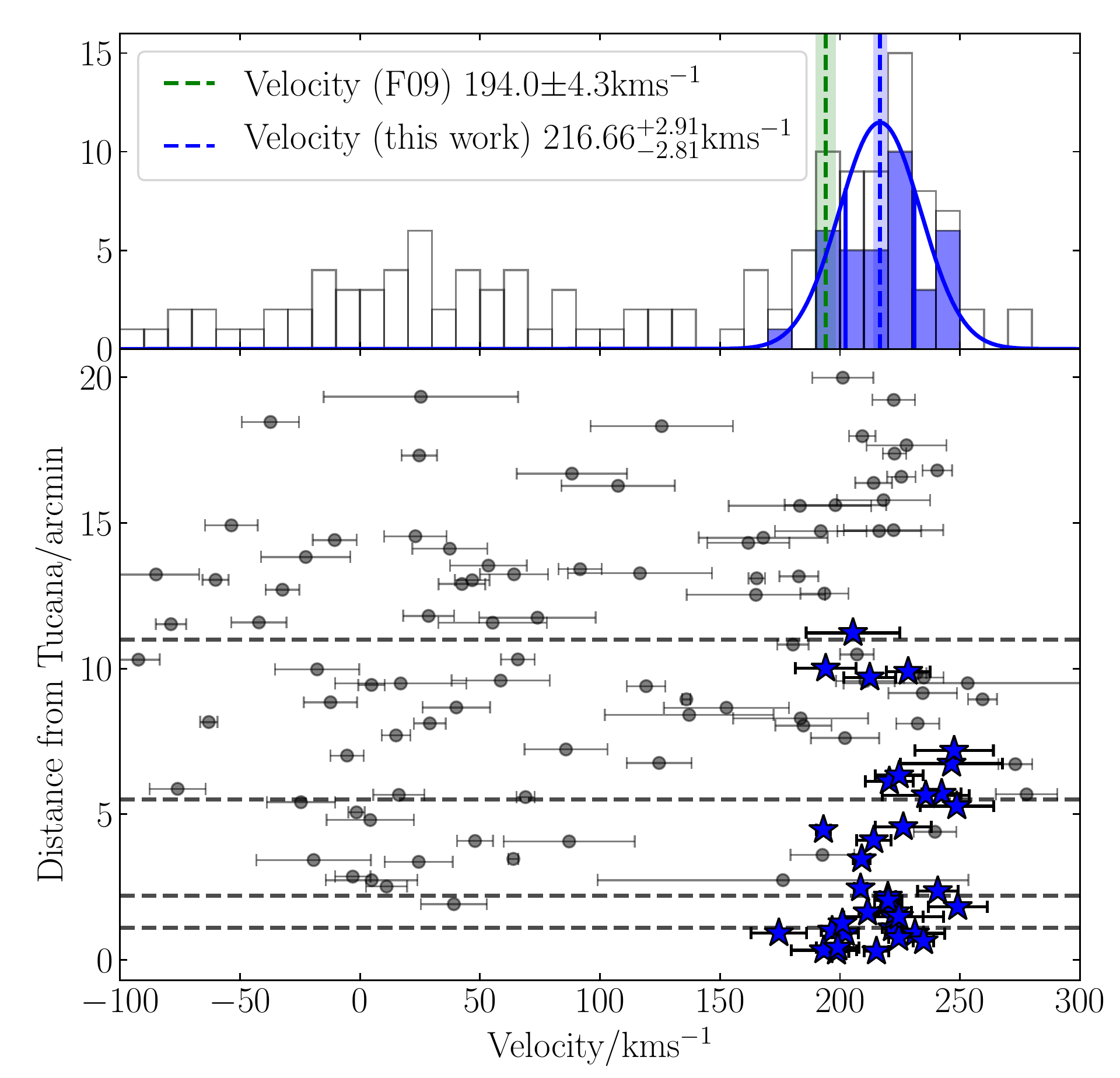}
	\caption{Top panel: Histogram of all measured velocities with members highlighted in blue, and the systemic velocity of Tucana as measured by this study and F09 marked as dashed lines. Bottom panel: Velocities of all observed objects plotted as a function of radius from the centre of Tucana, with the identified members marked as blue stars. The dashed lines mark 1$\times$, 2$\times$, 5$\times$ and 10$\times$ the half--light radius of Tucana.}
	\label{fig:velrad}
\end{figure}

The histogram of radial velocities of the identified Tucana members is shown in Fig. \ref{fig:velrad}. The systemic velocity of the peak has increased slightly, and is more than 3$\sigma$ outside of F09's value. It corresponds to a velocity relative to the Local Group of $v_{\mathrm{LG}}=+95.4$ kms$^{-1}$, confirming that the galaxy is receding from the Local Group.

\subsubsection{Comparison with Previous Study} \label{sec:f09comparison}
As previously noted, there is a significant offset between the systemic velocities of our FLAMES dataset and that measured by F09 of $\Delta v_{\mathrm{Tuc}}=22.7$ kms$^{-1}$. To investigate this, we tested the wavelength calibration of the spectra, which is provided by the initial data reduction pipeline, by cross correlating the raw spectra (before sky subtraction) with a reference sky template. This sky spectrum, which was produced during observations of the Sculptor dwarf galaxy by \citet{tolstoy04}, provides a reliable reference with which we can calibrate our data. Spectra with a velocity shift $>7$ kms$^{-1}$ were rejected as having a significant wavelength offset which could later affect our analysis of the radial velocity of the object. This process returned negligible shifts between the science and template spectra, confirming the validity of the wavelength calibration. We also rigorously tested the heliocentric correction applied to the spectra to ensure this was accurate. In a further test of the data, we randomly selected a sample of 20 stars (the size of the F09 sample) from our 36 members, and measured the systemic velocity of this sample. After 100 iterations, the difference between the sample velocity and the $v=194.0$ kms$^{-1}$ measured by F09 was $\Delta v>10$ kms$^{-1}$ in all cases, $\Delta v>20$ kms$^{-1}$ in 75 cases, and $\Delta v>22.7$ kms$^{-1}$ in 49 cases. The average offset was $\bar{\Delta v}=22.9$ kms$^{-1}$, consistent with the offset measured for the full dataset.

Comparing the velocities of the stars that overlap the two samples (see table \ref{tab:allobj}), we find some variation in the measurements but no systematic offset. Excluding object 100020 (21 in F09), which has an offset of 210 kms$^{-1}$ and is clearly indicative of a mismeasurement in one of the samples; in the overlapping objects we measure an average velocity offset of 11.3kms$^{-1}$. This value is similar to the average velocity error of our identified members ($\bar{{v}_{\mathrm{err}}}=9.3$ kms$^{-1}$) but is somewhat smaller than the offset between the measured systemic velocities. We also note that the 3 non members listed in Table 3 of F09 are matched to non members in the FLAMES data. Many of the objects display an offset $>20$ kms$^{-1}$; however all but one (with the exception of object 100020) have velocities consistent within 3--$\sigma$.

Therefore, whilst the offset appears significant, we believe our result for the velocity can be considered reliable, especially given the larger sample size relative to F09. Furthermore, the offset does not affect our measurement of the velocity dispersion nor any conclusions drawn from this.

\subsubsection{Mass of Tucana}\label{sec:tucmass}

A number of estimators have been defined to determine the virial mass of a dwarf galaxy from its velocity dispersion. These estimators take the general form

\begin{eqnarray}
    M\mathrm{_{est}}(<\lambda\rm{R})=\frac{\mu r\mathrm{_{half}} \sigma_{v,\mathrm{half}}^{2}}{G},
\label{eq:mass}
\end{eqnarray}

\noindent where $r_{\mathrm{half}}$ is the half--light radius of the galaxy, and $\sigma_{v,\mathrm{half}}$ is the velocity dispersion at that radius. Assuming a flat velocity dispersion profile, \citet{walker09} define $\lambda$=1 and $\mu$=3.5, such that the mass estimate becomes

\begin{eqnarray}
    M(<r_{\rm{half}})=580 r\mathrm{_{half}} \sigma_{v,\mathrm{half}}^{2}
\label{eq:walkermass}
\end{eqnarray}

\noindent From this, we determine the mass of Tucana to be $M_{\mathrm{half}}=3.4^{+1.5}_{-1.3}\times10^{7}$ M$_{\odot}$, corresponding to a mass--to--light ratio within the half--light radius of $M/L\mathrm{_{half}}$$\approx62^{+27}_{-23}$  M$_{\odot}$/L$_{\odot}$, assuming a luminosity for Tucana of $5.5\times10^{5}$ L$_{\odot}$. This result is slightly lower than the estimated mass quoted in F09, who found a pressure supported total mass of $M\approx5.0\times10^{7}$ M$_{\odot}$. However, this was determined by assuming that mass follows light in the system, and using an estimator to determine the mass to light ratio, from which the mass is calculated. Given that Tucana is a centrally dense galaxy which is likely to be highly dark matter dominated, this method is less accurate than equation \ref{eq:walkermass}, which derives the mass directly from the measured velocity dispersion. Using the \citet{walker09} method with the velocity dispersion calculated by F09 returns a value of $M_{\mathrm{half}}=4.1^{+2.3}_{-1.8}\times10^{7}$ M$_{\odot}$, which is consistent with our result.

Recently, \citet{errani18} redefined the mass estimator with $\lambda$=1.8 and $\mu$=3.5. This does not require a flat velocity dispersion profile, and so is insensitive to any fluctuations in the profile (see e.g. Fig. \ref{fig:gravsphere_result}). Using this new estimator, we determine a mass for Tucana of $M_{\mathrm{half}}=8.6^{+3.7}_{-3.2}\times10^{7}$ M$_{\odot}$, corresponding to a mass--to--light ratio within the half--light radius of $M/L\mathrm{_{half}}\approx156^{+68}_{-58}$ M$_{\odot}$/L$_{\odot}$. This is consistent with the result of F09, though it is not directly comparable to the Walker estimate as the enclosed radius is larger.

\subsection{Rotation in Tucana}

\begin{figure}
	\centering
	\includegraphics[width=\linewidth] {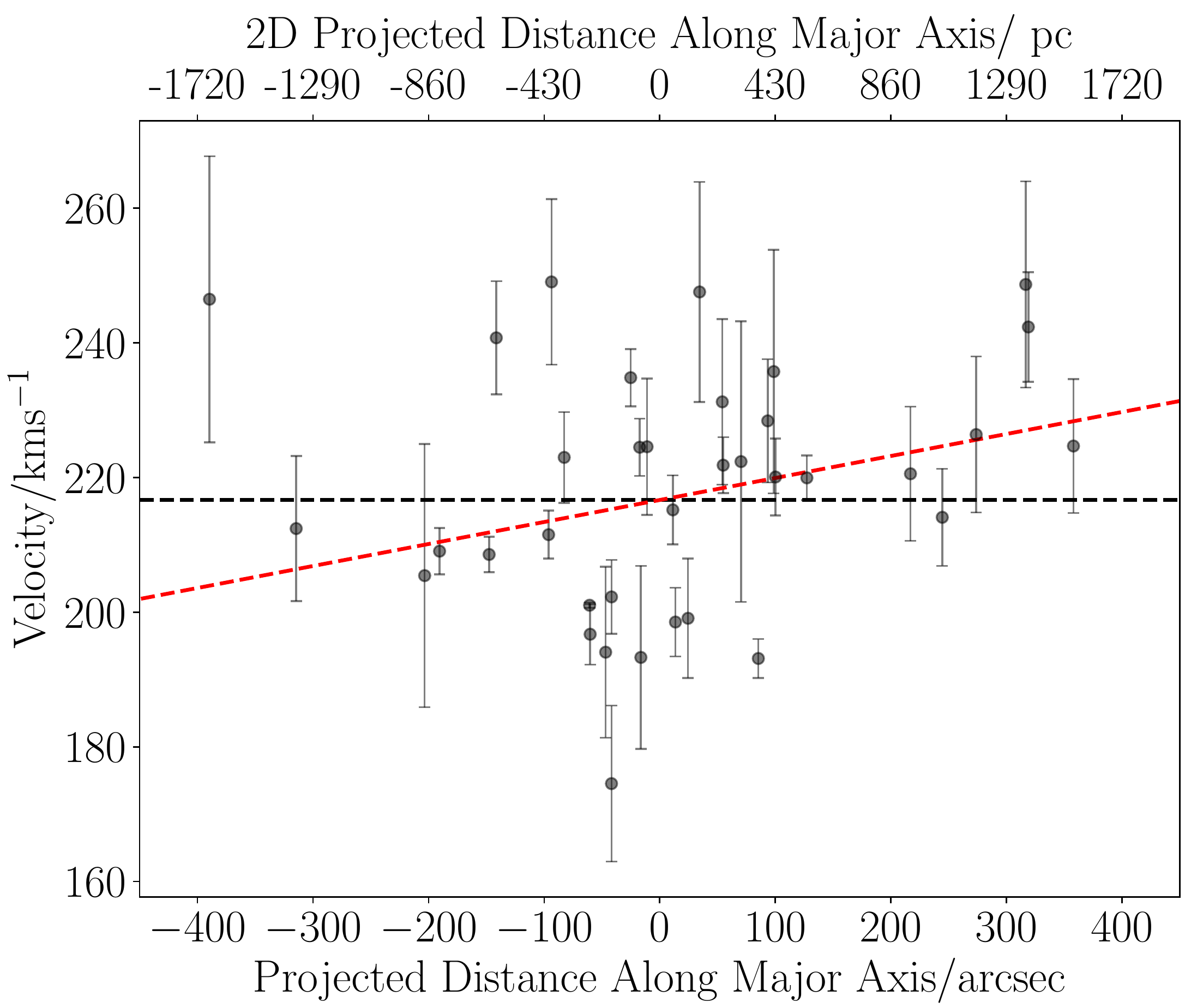}
	\caption{Radial velocities of the 36 identified Tucana members as a function of their projected position along the major axis. The red dashed line highlights the measured velocity gradient of $\frac{dv_{r}}{d\chi}=7.6^{+4.2}_{-4.3}$ kms$^{-1}$ kpc$^{-1}$. The grey dashed line shows the measured systemic velocity of the system.}
	\label{fig:velmajorax}
\end{figure}

F09 find a rotation signature for Tucana of $v_{\mathrm{rot}}=16$ kms$^{-1}$, by fitting a rotation curve to the radial velocities of their member stars. To search for this in our data, we plot the radial velocities of our member stars as a function of projected distance along the major axis in Fig. \ref{fig:velmajorax}. This highlights a possible gradient across the data. We quantify this gradient by utilising another \texttt{emcee} routine. \citet{martin10} redefine the Gaussian representing the Tucana peak as

\begin{eqnarray}
P_{\mathrm{Tuc}}=\frac{1}{\sqrt{2\pi}\sqrt{\sigma_{v, \mathrm{Tuc}}^2+v_{\mathrm{err}, i}^2}} \times  \exp\bigg(-\frac{1}{2}\bigg[\frac{\Delta v_{r,i}^{2}}{\sqrt{\sigma_{v, \mathrm{Tuc}}^2+v_{\mathrm{err}, i}^2}}\bigg]^2\bigg),
\label{eq:gradient}
\end{eqnarray}

\noindent where $\Delta v_{r,i}$ is the difference between the velocity of a star and a velocity gradient $\frac{dv_{r}}{d\chi}$ acting along the angular distance of a star along an axis $y_i$ with position angle $\theta$, as shown in equation \ref{eq:deltav}.

\begin{eqnarray}
\Delta v_{r,i}=v_{r,i}-\frac{dv_{r}}{d\chi}y_{i}+\bar{v_{r}}
\label{eq:deltav}
\end{eqnarray}

\noindent $y_i$ can be determined from the RA and Dec. of the star, ($\alpha_{i}$, $\delta_{i}$), and of the centre of Tucana, ($\alpha_{0}$, $\delta_{0}$), using 

\begin{eqnarray}
y_i&=&X_{i}\sin{\theta}+Y_{i}\cos{\theta}\\
X_i&=&(\alpha_{i}-\alpha_{0})\cos(\delta_{0})\\
Y_i&=&\delta_{i}-\delta_{0}.
\label{eq:gradeqns}
\end{eqnarray}

\noindent By replacing equation \ref{eq:probmw} with equation \ref{eq:gradient} in the MCMC routine, we can generate a fit to the data which accounts for the velocity gradient $\frac{dv_{r}}{d\chi}$ produced by rotation. We introduce flat priors for the new parameter $\frac{dv_{r}}{d\chi}$ such that $-40<\frac{dv_{r}}{d\chi}<40$. If $\theta$ is fixed to match the position angle of the major axis ($\theta=97^{\circ}$), we measure a rotation gradient of $\frac{dv_{r}}{d\chi}=7.6^{+4.2}_{-4.3}$ kms$^{-1}$ kpc$^{-1}$, with the systemic velocity measured as $v_{\mathrm{Tuc}}=215.2_{-2.7}^{+2.8}$ kms$^{-1}$ and the velocity dispersion as $\sigma_{v,\mathrm{Tuc}}=13.3_{-2.3}^{+2.7}$ kms$^{-1}$. This rotation gradient, marked as the red dashed line in Fig. \ref{fig:velmajorax}, equates to a rotation velocity of $2.2\pm1.2$kms$^{-1}$ at the half light radius. If $\theta$ is allowed to evolve freely (with a flat prior of $0<\theta<\pi$), we determine the best values to be: $v_{\mathrm{Tuc}}=214.9_{-3.2}^{+3.2}$ kms$^{-1}$; $\sigma_{v,\mathrm{Tuc}}=13.5_{-2.3}^{+2.8}$ kms$^{-1}$; $\frac{dv_{r}}{d\chi}=6.1^{+4.6}_{-4.8}$ kms$^{-1}$kpc$^{-1}$; $\theta=86.5^{\circ}$$^{+37.8}_{-35.5}$; amounting to a rotation velocity of $1.7\pm1.3$kms$^{-1}$ at the half light radius.

It therefore appears that there is a small velocity gradient in Tucana due to the presence of rotation, consistent with alignment with the major axis. The mass estimator from \citet{walker09} (equation \ref{eq:mass}) assumes a velocity dispersion dominated system. If we recalculate the dynamical mass using the smaller velocity dispersion accounting for rotation, we obtain a mass of $M_{\mathrm{half}}=2.9^{+1.3}_{-0.9}\times10^{7}$ M$_{\odot}$, corresponding to a mass--to--light ratio within the half--light radius of $M/L\mathrm{_{half}}$$\approx53^{+24}_{-17}$  M$_{\odot}$/L$_{\odot}$. These results are within the uncertainty ranges of the non--rotating result, and given that the rotation gradient is small relative to the velocity dispersion, Tucana is still classed as a dispersion dominated galaxy.

\section{Modelling the Dark Matter Density Profile}\label{sec:gravsphere}

The high velocity dispersion of Tucana suggests that it has a high central dark matter density. In the absence of star formation, a steep central cusp is predicted to be present in the density profiles of all galaxies \citep{navarro96}, but thus far the majority of observations of dSphs show a slight preference towards flattened cores (see the discussion in section \ref{sec:intro}). The dearth of recent star formation in Tucana make it a strong candidate for hosting a `pristine' cusp \citep{brook15,read18a,bermejo18}. In this section, we perform Jeans modelling of the data for Tucana to estimate its central dark matter density and quantitatively test this idea.

\begin{figure*}
	\centering
	\begin{minipage}{0.348\textwidth}
	\includegraphics[width=\linewidth]{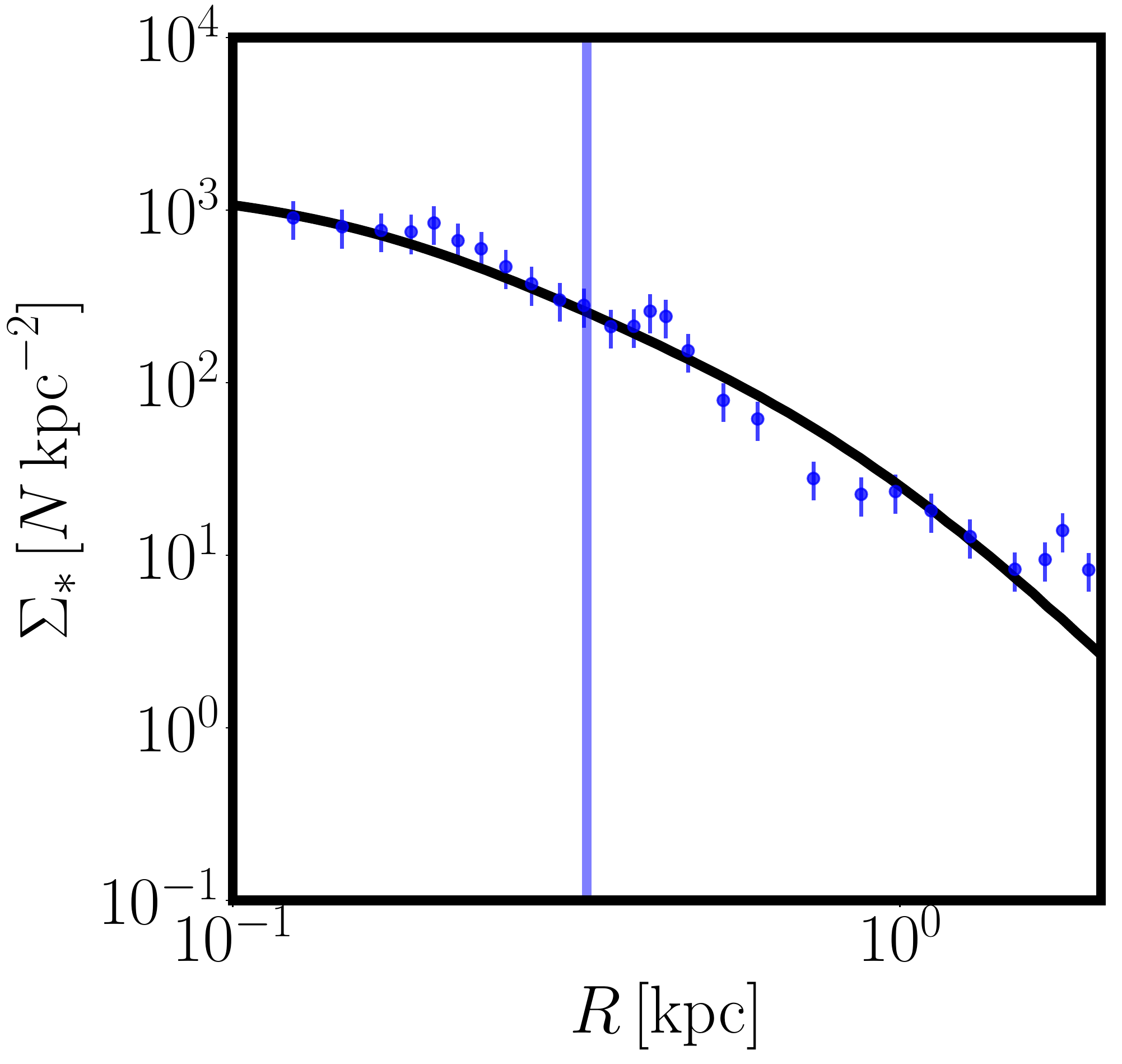}
	\end{minipage}
	\hspace{15mm}
	\begin{minipage}{0.33\textwidth}
	\includegraphics[width=\linewidth]{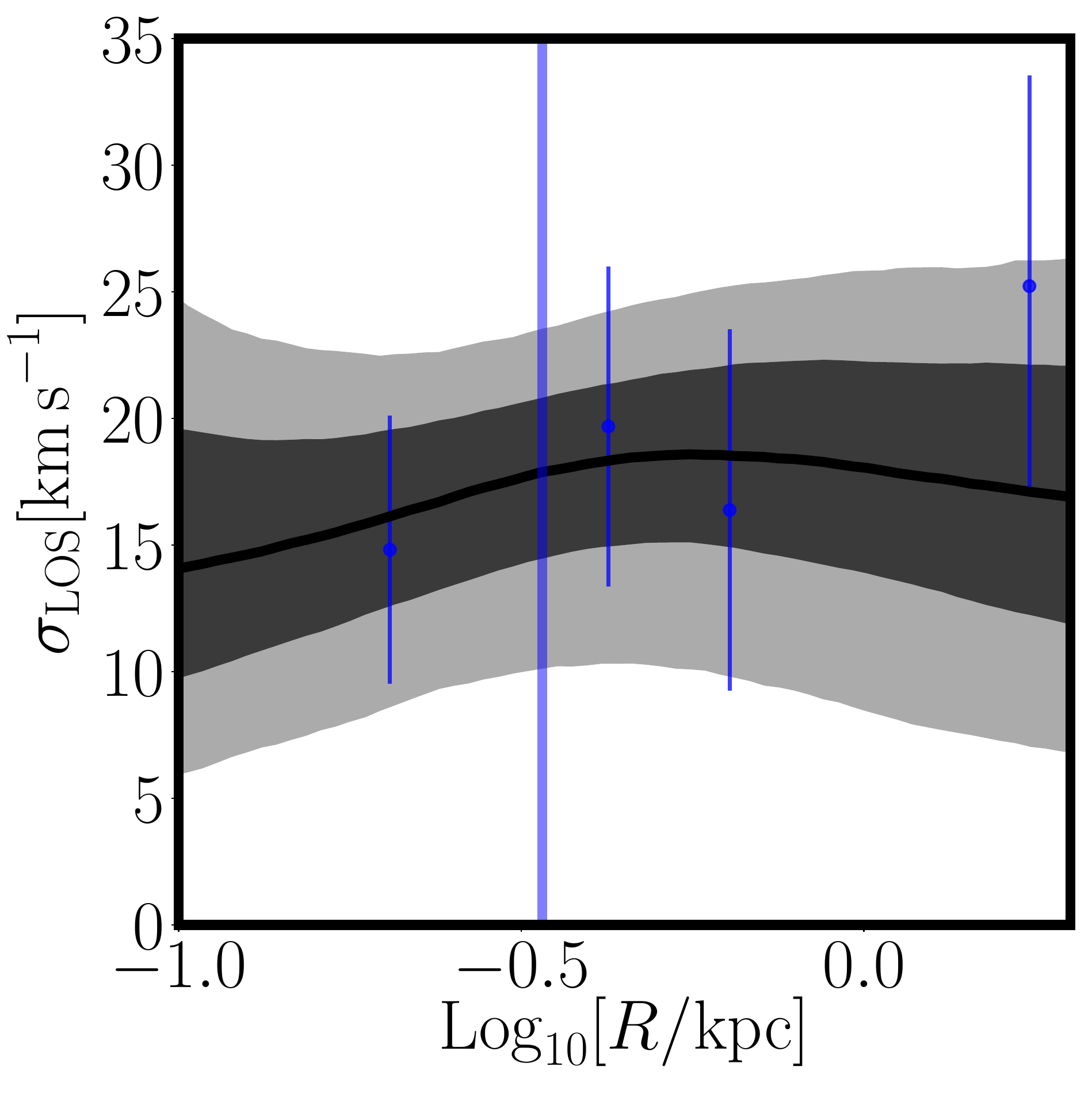}
	\end{minipage}
	\caption{\textsc{GravSphere} model fits to the surface density profile, left, and line--of--sight velocity dispersion profile, right, for Tucana. The blue points mark the input `observed' data; the black line is the fitted profile, with grey contours showing 1-- and 2--$\sigma$ uncertainty ranges. The vertical line marks the half--light radius of Tucana as modelled by \textsc{GravSphere}.}
	\label{fig:gravsphere_models}
\end{figure*}

\subsection{The \textsc{GravSphere} Mass Modelling Code}

\citet{read17b} introduced a new non--parametric Jeans code \textsc{GravSphere}, which returns the density profile $\rho$(r) and velocity anisotropy $\beta$(r) of a system using only line--of--sight velocities (see also \citealt{read18a}). \textsc{GravSphere} solves the Jeans equation \citep{jeans1922} for a set of `tracers' (i.e. stellar members) of a spherical mass distribution defined by the radial density profile $\rho$(r) and velocity anisotropy $\beta$(r). The projected spherical Jeans equation is given by \citet{binney82} as 

\begin{eqnarray}
\sigma^{2}_{\mathrm{LOS}}(R)=\frac{2}{\Sigma(R)}\int_{R}^{\infty}\bigg(1-\beta\frac{R^{2}}{r^{2}}\bigg)v\sigma^{2}_{r}\frac{r \mathrm{d}r}{\sqrt{r^{2}-R^{2}}},
\label{eq:jeans}
\end{eqnarray}

\noindent where $\Sigma(R)$ is the surface mass profile at projected radius R, $v(R)$ is the spherically averaged tracer density, and $\beta$(r) is the velocity anisotropy,

\begin{eqnarray}
\beta=1-\frac{\sigma_t^2}{\sigma_r^2}.
\label{eq:beta}
\end{eqnarray}

\noindent $\sigma_t$ and $\sigma_r$ are the tangential and radial velocity dispersion profiles. $\sigma_r$ is given by \citep{vandermarel94,mamon05}

\begin{eqnarray}
\sigma_r^2(r)&=&\frac{1}{v(r) g(r)}\int_r^\infty\frac{G M(\tilde{r}) v(\tilde{r})}{\tilde{r}^2} g(\tilde{r}) \mathrm{d}\tilde{r}\\
g(r)&=&\exp\bigg(2\int \frac{\beta(r)}{r}\mathrm{d}r \bigg).
\label{eq:sigr}
\end{eqnarray}

\noindent $M(r)$ is the cumulative mass of the galaxy. \textsc{GravSphere} uses a non--parametric model for $M(r)$, consisting of a contribution from all visible matter, and a contribution from dark matter modelled by a series of power laws centred on a set of radial bins. The tracer light profile uses a series sum of Plummer spheres \citep{plummer11} and so is also non--parametric. The code fits this model to the surface density profile $\Sigma_{*}(R)$ and line--of--sight velocity dispersion profile $\sigma_{\mathrm{LOS}}(R)$ using \texttt{emcee} \citep{foremanmackey13}. To avoid infinities in $\beta$, a symmetrised version of $\beta$ is used in the model. $\tilde{\beta}$ is defined in equation \ref{eq:betastar}, and describes the distribution of velocities in the system, with $\tilde{\beta}=0$ describing an isotropic velocity distribution, $\tilde{\beta}=1$ a fully radial distribution and $\tilde{\beta}=-1$ a fully tangential distribution \citep{read06b}.

\begin{eqnarray}
\tilde{\beta}=\frac{\sigma_r^2-\sigma_t^2}{\sigma_r^2+\sigma_t^2}=\frac{\beta}{2-\beta}.
\label{eq:betastar}
\end{eqnarray}

\noindent $\tilde{\beta}$ is given a flat prior of $-1<\tilde{\beta}<1$, to give equal weight to fully radial and fully tangential distributions. \textsc{GravSphere} also uses virial shape parameters to obtain constraints on $\beta$ using only line--of--sight velocities, thus breaking the well--known $\rho$--$\beta$ degeneracy \citep{binney82,merrifield90,read17b}. The code can be used to determine $\rho$(r) and $\beta$(r) for any near--spherical stellar system, such as star clusters, spheroidal or elliptical galaxies and galaxy clusters. \textsc{GravSphere} has been extensively tested on mock data (\citealt{read17b}, and see below). For a more complete description of \textsc{GravSphere}, we refer the reader to \citet{read17b}.

In addition to the `free--form' dark matter model described above, GravSphere can also fit the cosmologically--motivated \textsc{coreNFW} profile. This profile was originally designed to fit simulations of halos that have undergone `dark matter heating' \citep{read16}, but also provides a good fit to dark matter halos in a self--interacting dark matter cosmology \citep{read18a}. It has the advantage that it fits a dark matter core size parameter that can be connected to a self--interaction cross section for dark matter \cite{read18a}, and a halo virial mass, $M_{200}$, that can be compared with cosmological expectations from abundance matching (e.g. \citealt{read18c}). We present the results of fitting this \textsc{coreNFW} model to Tucana in appendix \ref{appendixa}.

\textsc{GravSphere} has been extensively tested on mock data, including mocks that break spherical symmetry, that include foreground contamination and binary stars, and that are being tidally stripped by a larger host galaxy \citep{read17b,read18a}. However, all tests to date have focussed on mocks with $>$500 member velocities, which is an order of magnitude more than we have available for Tucana. For this reason, we present additional mock data tests in appendix \ref{appendixb}. These are set up to mimic Tucana, modelling the selection function, contamination and sampling similarly to our real Tucana data. We show that we are able to correctly infer the dark matter density at 150pc from the centre of Tucana with kinematics for just 36 member stars, albeit with substantially larger uncertainties than we obtain for data with 500 member stars. Similarly to the findings in \citet{read18a}, we find that \textsc{GravSphere} is not able to distinguish cusps from cores with this many member stars. However, as explained in \citet{read18a,read18b}, an inference of the central dark matter density, $\rho_{\rm DM}(150\,{\rm pc})$, is sufficient to constrain interesting dark matter models. For this reason, we focus in this paper on obtaining an estimate of $\rho_{\rm DM}(150\,{\rm pc})$ for Tucana.

\subsection{Dark Matter Density Profile}

\begin{figure}
	\centering
	\includegraphics[width=\linewidth]{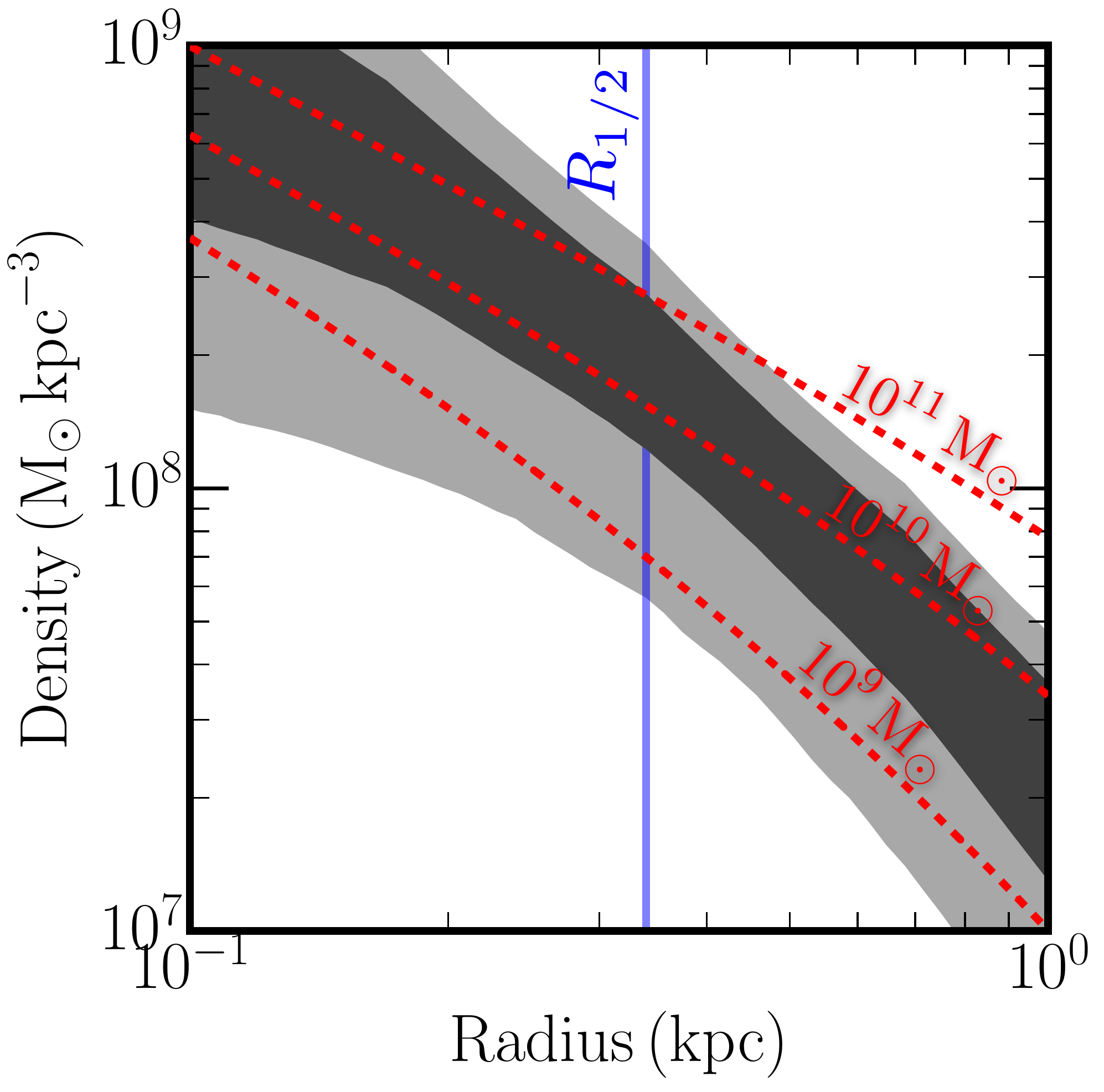}
	\caption{The density profile of Tucana generated by \textsc{GravSphere}. Grey contours show 1-- and 2--$\sigma$ uncertainty ranges, and the vertical line marks the half--light radius of Tucana as modelled by \textsc{GravSphere}. Dashed red lines mark NFW profiles of different pre--infall halo masses. The data weakly favour the $10^{10}$ M$_{\odot}$ model, while the central density $\rho_{\rm DM}(150\,{\rm pc}) > 10^8$\,M$_\odot$\,kpc$^{-3}$ at better than 2-sigma confidence, consistent with a dark matter cusp.}
	\label{fig:gravsphere_result}
\end{figure}

In this section, we use \textsc{GravSphere} to estimate the dark matter density profile of Tucana. For this, we require the surface brightness profile, $\Sigma_*(R)$, and velocity dispersion profile, $\sigma_{\rm{LOS}}(R)$, of our member stars. The former was generated from RGB stars in the Magellan/ MegaCam photometry, adjusted to correct for the fact that our observed fields are misaligned with the central coordinates of Tucana. The velocity dispersion profile was modelled using the radial velocities we have obtained with FLAMES+GIRAFFE. To generate the velocity dispersion profile, the probabilities of membership of our stellar sample are summed, resulting in a membership probability weighted number of stars $N_{\rm{eff}}=\sum_{i=1}^{N_{\rm{mem}}}P_{\rm{mem,i}}$. For our 36--strong sample, this equates to $\sim$20 effective members. The data is binned by radius from the centre of Tucana, with each bin containing the same membership weighted number of stars. The $\Sigma_{*}(R)$ and $\sigma_{\mathrm{LOS}}(R)$ profiles are shown in Fig. \ref{fig:gravsphere_models}. Note that the vertical blue line on these plots marks the half--light radius, $r_{\mathrm{half}}=340$ pc, as modelled from the surface brightness profile. This is just outside 1--$\sigma$ larger than the literature value (see table \ref{tab:tuc_properties}).

We model Tucana with \textsc{GravSphere} under the assumption that it is a spherical, non rotating system. Although we detect a small rotation gradient in Tucana, its effect on the velocity dispersion measurement is negligible with respect to the size of the uncertainties, and so it has no effect on the \textsc{GravSphere} model. We use a stellar mass for Tucana of M$_*$=3.2$\times10^6 M_{\odot}$ \citep{hidalgo13}, assuming an error on M$_*$ of 25$\%$. The final density profile is shown in Fig. \ref{fig:gravsphere_result}. The red dashed lines highlight NFW profiles at different masses. These were produced using the concentration--mass relation from \citet{dutton14}, but multiplied by 1.4 to account for the fact that subhalos in the Aquarius simulations are found to be systematically more concentrated than field halos \citep{springel08}.

The profile favours a high central density of $\rho_{\mathrm{DM}}$(150 pc)$=5.5_{-2.5}^{+3.2} \times 10^8$ M$_{\odot}$ kpc$^{-3}$, suggesting that Tucana is as dense, if not more so, than Draco \citep{read18a}. The profile appears consistent with the presence of a pristine cusp within 1$\sigma$. Using the \textsc{core}NFW fit presented in appendix \ref{appendixa}, we also infer a pre--infall halo mass of $M_{200}=1.37^{+0.49}_{-0.44} \times 10^{10}$ M$_{\odot}$ for Tucana, consistent with the findings of \citet{brook15}.

\section{Discussion}\label{sec:disc}

\subsection{Comparison to Previous Work}

\begin{figure*}
	\centering
	\includegraphics [width=0.9\textwidth] {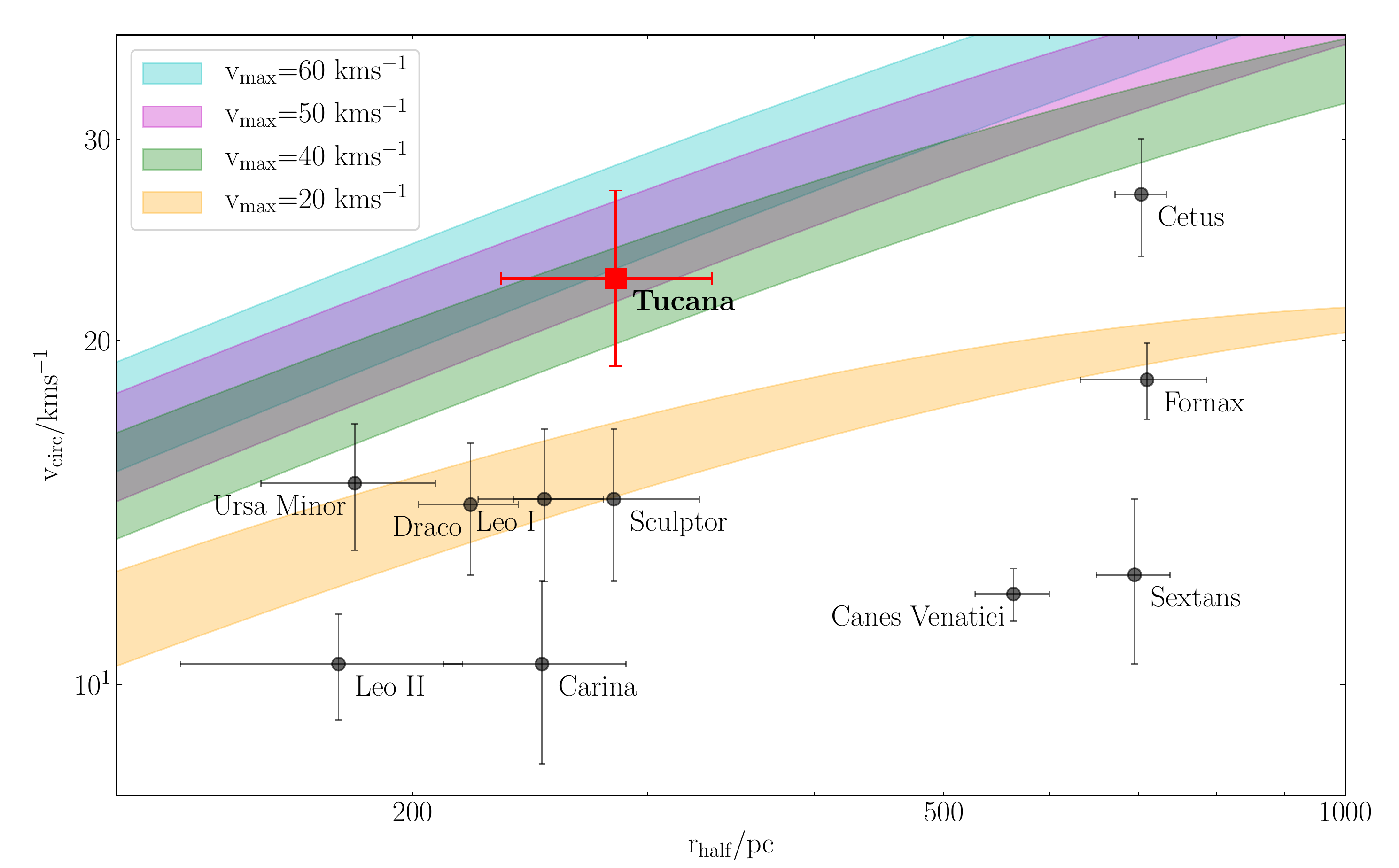}
	\caption{Circular velocities (and errors) of Local Group dwarf galaxies plotted as a function of half--light radius, with the results of this study shown in red. Rotation curves corresponding to NFW profiles with $v_{\mathrm{max}}=(20,40,50,60)$ kms$^{-1}$ are highlighted. Tucana clearly resides within a `massive failure' subhalo with $v\mathrm{_{max}}=40$ kms$^{-1}$. Dwarf galaxy data taken from \citet{mcconnachie12}.}
	\label{fig:tbtf_plot}
\end{figure*}

Our velocity dispersion of $\sigma_{\mathrm{v,Tuc}}=14.4_{-2.3}^{+2.8}$kms$^{-1}$ is remarkably similar to that measured by F09, with a difference of just $\sim1$ kms$^{-1}$. It should be noted that the $\sigma_{v,\mathrm{Tuc}}=15.8$ kms$^{-1}$ quoted by F09 accounts for a potential rotation in signature Tucana of 16 kms$^{-1}$. Our non--rotational dispersion result is within 1$\sigma$ of their non--rotational value of $\sigma_{v,\mathrm{Tuc}}=17.4^{+4.5}_{-3.5}$ kms$^{-1}$, so is also consistent with the like--for--like result. We detect a slightly smaller rotation signature of $\frac{dv_{r}}{d\chi}=7.6^{+4.2}_{-4.3}$ kms$^{-1}$.

Fig. \ref{fig:velrad} plots the velocities of the observed stars as a function of radius from the centre of Tucana, with those identified as members highlighted in blue. All members are shown to lie within $\sim10 r\mathrm{_{half}}$ of Tucana. There is a selection bias in our observation method; by using a fibre spectrograph, we preferentially observe more distant members, as they are less closely packed,  compared to those in the dense centre. Indeed, there is a large population of stars with velocities close to the systemic velocity of Tucana outside $10 r\mathrm{_{half}}$. We have defined our probability functions such that these objects are not selected as members due to their large distances.

There is a substantial velocity offset between the systemic velocity quoted by F09 and that measured from our data of around $\Delta v_{\mathrm{Tuc}}$ $\approx23$ kms$^{-1}$. This could be due to a number of factors, such as a wavelength miscalibration or a bias in the cross correlation procedure used. We also note that as a low resolution slit--based spectrograph, FORS2 (as used in F09) is not optimised for velocity measurements and may suffer from variations in the velocity zero point, as discussed in \citet{kacharov17}'s study of the Phoenix dwarf galaxy. Without repeat measurements from FLAMES+GIRAFFE, we cannot conclusively identify the origin of the velocity offset. However, we have thoroughly tested our calibration, as described in section \ref{sec:f09comparison}, and have confidence in our result. It also should be noted that the offset does not influence our velocity dispersion measurement or any ensuing conclusions. Despite the offset, our result supports the conclusion of F09 that the galaxy is receding from the Milky Way, and could continue to do so, unbound from the Local Group.

\subsection{Tucana: A Massive Failure}

Our velocity dispersion result suggests a high central density consistent with Tucana residing within a supposed `massive failure' halo as predicted to exist by \citet{boylankolchin11}. This would make Tucana the first known exception to the `too--big--to--fail' problem, whereby simulated subhalos are too centrally dense to host the observed dwarf galaxies \citep{boylankolchin11,boylankolchin12}. In \citet{wang12}, too--big--to--fail is restated in terms of the galaxies' maximum circular velocities; all known satellites are observed with $v_{\mathrm{max}}<30$ kms$^{-1}$ (with the exception of the Magellanic Clouds--- \citealt{jiang15}), yet the simulated halos should host galaxies with 40 kms$^{-1}<v_{\mathrm{max}}<60$ kms$^{-1}$. Our measured velocity dispersion corresponds to a circular velocity for Tucana of $v_{\mathrm{circ}}=22.7_{-4.7}^{+5.4}$ kms$^{-1}$. We plot this velocity alongside those of the nine bright Local Group dSphs (those plotted in \citealt{boylankolchin12}, Fig. 1), plus the isolated dSph Cetus, as a function of half--light radius in Fig. \ref{fig:tbtf_plot}. The circular velocity of Tucana is significantly higher than that of other Local Group dwarf galaxies, indicating that Tucana is much more centrally dense than the typical dSph. Plotting rotation curves for different maximum circular velocities suggests that Tucana resides in a halo with maximum velocity $v_{\mathrm{max}}>40$ kms$^{-1}$. In other words, Tucana appears to reside in a `massive failure' halo. If this is the case, Tucana would be the only dwarf spheroidal galaxy in the Local Group known to reside in such a halo.

\begin{figure}
	\centering
	\includegraphics [width=\linewidth] {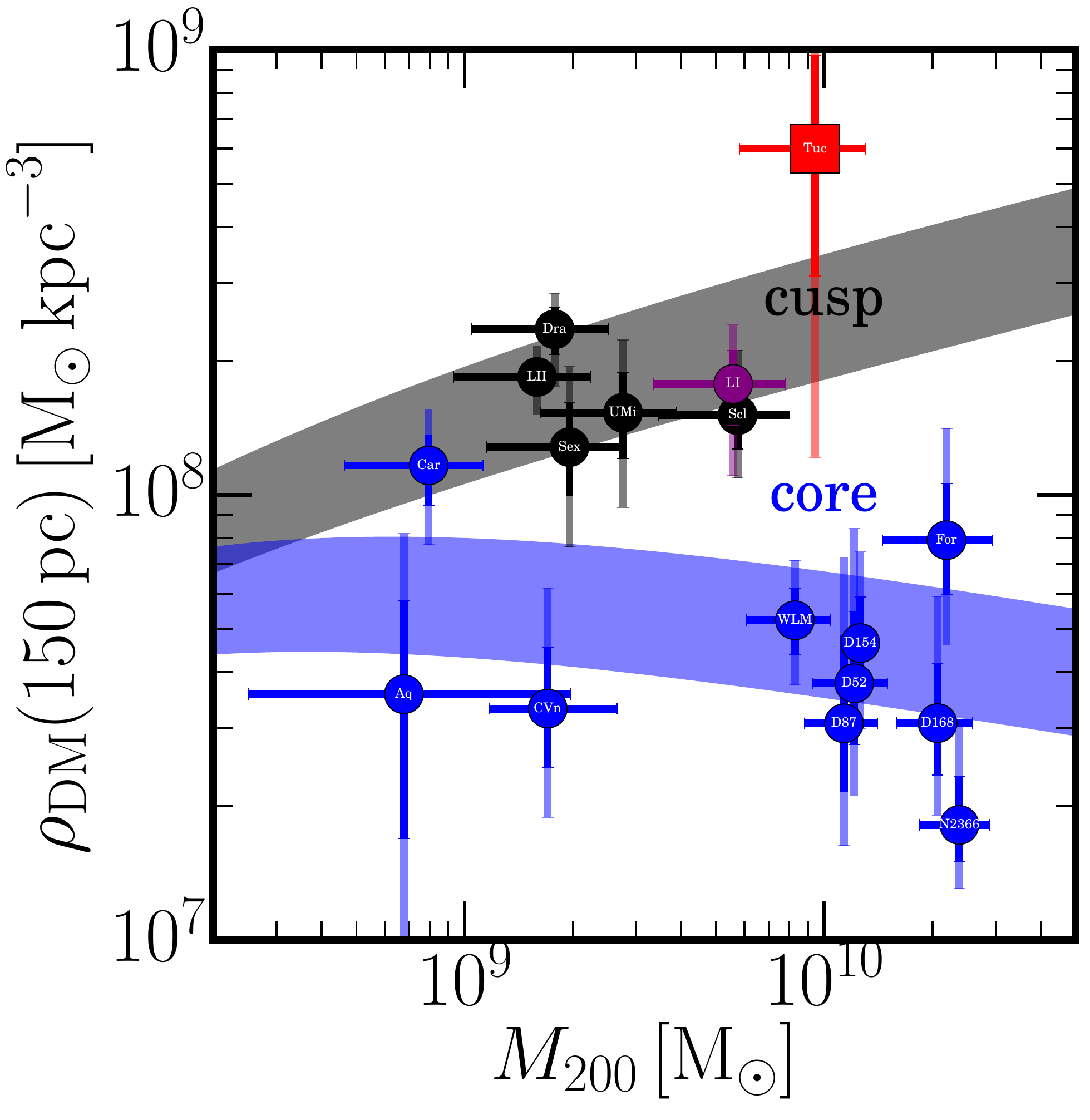}
	\caption{Inner DM density as a function of pre--infall halo mass for a selection of Local Group dwarf galaxies, as modelled in \citet{read18b}. Black points shows galaxies whose star formation was truncated more than 6 Gyrs ago; blue points show galaxies truncated less than 3 Gyrs ago; and the purple points were truncated between 3 and 6 Gyr ago. The shaded regions highlight the expected DM density of cusped and cored profiles. Tucana is shown as the red point. In addition to highlighting the correlation between star formation history and the shape of the density profile, this plot shows the high central DM density of Tucana relative to other Local Group dwarf galaxies, a result which appears to be consistent with a central cusp.}
	\label{fig:dm_heating}
\end{figure}

The too--big--to--fail problem has been widely viewed as a universal problem with the $\Lambda$CDM model. Originally discovered to be an issue in the Milky Way satellite population, both \citet{collins14} and \citet{tollerud14} find that the problem is also present in M31 satellites. \citet{read17a} show that both too--big--to--fail and `missing satellites' are confined to group environments, and so must be the result of galaxy formation physics as opposed to some exotic formulation of the cosmology. Our result shows that there are conditions under which galaxies can retain their central mass and reside in the $\Lambda$CDM predicted halos. Tucana's isolation from tidal effects and quenched star formation history mean it has been unaffected by the baryonic feedback effects usually invoked to resolve too--big--to--fail. This would make Tucana unique even among isolated dwarfs: \citet{kirby14} find that star formation in other isolated Local Group galaxies has been energetic enough to lower the central density, such that these galaxies are fully consistent with Milky Way/ M31 satellites. However, the shut--down in star formation in Tucana around 10 Gyr ago \citep{monelli10} likely allowed it to retain its central mass. 

\subsection{A Pristine Cusp in Tucana}

Fig. \ref{fig:dm_heating} shows the central DM density for a range of Local Group dwarfs (the gas rich ones were not modelled with GravSphere) as a function of pre--infall halo mass \citep{read18b}. The figure highlights a strong correlation between star formation history and inner DM density, with galaxies whose star formation shut down long ago possessing central densities consistent with a central cusp, and those with extended star formation histories consistent with a cored profile. As described in \citet{read18b}, this supports the hypothesis that dark matter is heated up by extended, bursty star formation, reducing the central density and transforming cusps to cores.

Using the abundance matching technique from \citet{read18c}, with a stellar mass of M$_*=3.2\times10^6$ M$_{\odot}$ \citep{hidalgo13} and the star formation history for Tucana from \citet{monelli10}, we estimate a halo mass for Tucana of M$_{200, \rm{abund}}=9.4 \pm 3.6\times10^9$ M$_{\odot}$, in excellent agreement with our \textsc{GravSphere} models (see appendix \ref{appendixa}). To be consistent with the results presented in \citet{read18b}, we use M$_{200, \rm{abund}}$ and $\rho_{DM}$(150 pc) as inferred using \textsc{GravSphere} to plot Tucana on Fig. \ref{fig:dm_heating} (red data point). The high central density returned by our \textsc{GravSphere} modelling places Tucana above the other dwarf galaxies in this plot, and is consistent with the expectations of a cusped profile within 1$\sigma$. This result is again consistent with the limited period of star formation experienced by Tucana.

The errors on the inner DM density are very large. However, Tucana is so dense (i.e. its velocity dispersion is so high) that we can be confident that it is more dense than WLM, Fornax and the other isolated dIrrs at 150pc at better than 95$\%$ confidence. Tucana is consistent with expectations for the inner density of `pristine' DM halos in $\Lambda$CDM that have undergone no DM heating, as expected from Tucana's old stellar population. Based on abundance matching and stellar kinematics, Tucana has a total halo mass consistent with WLM ($M_{200}=0.83\pm 0.2 \times 10^{10}$ M$_{\odot}$, \citealt{read17a}). WLM's inner rotation curve favours a central dark matter core over a cusp and is substantially less dense than what we find for Tucana in this work, as shown in Fig. \ref{fig:dm_heating}. This is because unlike WLM, Tucana's star formation ceased after just $\sim$1-2 Gyr --most likely due to ram pressure stripping by the Milky Way (e.g. \citealt{teyssier12,gatto13}). Our results are consistent with the idea that Tucana is much denser than WLM because it had insufficient star formation to undergo significant DM heating.

\section{Conclusions} \label{sec:conclusion}
Using the GIRAFFE spectrograph, we have taken high resolution spectra of the Tucana dwarf galaxy, identifying 36 member stars, and used a cross correlation method to measure their radial velocities. We make several key findings:

\begin{itemize}

\item We find the systemic velocity of Tucana to be $v_{\mathrm{Tuc}}=216.7_{-2.8}^{+2.9}$kms$^{-1}$, corresponding to $v_{\mathrm{GSR}}=121.7_{-2.8}^{+2.9}$kms$^{-1}$. This velocity is receding from the Local Group, and is consistent with the conclusion of F09 that Tucana has long been an isolated dwarf galaxy, which may have interacted with the Milky Way some 10 Gyr ago.

\item We measure a rotation gradient across Tucana of $\frac{dv_{r}}{d\chi}=7.6^{+4.2}_{-4.3}$ kms$^{-1}$ kpc$^{-1}$, which equates to a rotation velocity of $2.2\pm1.2$kms$^{-1}$ at the half light radius. The rotation appears to be aligned with the major axis of Tucana. Despite this rotation signature, Tucana is still primarily a dispersion supported system.

\item The velocity dispersion of Tucana is found to be $\sigma_{\mathrm{v,Tuc}}=14.4_{-2.3}^{+2.8}$kms$^{-1}$. This dispersion is consistent with the result of F09, and suggests that Tucana is significantly more centrally dense than other dSphs. Tucana is found to be compatible with high density subhalos predicted by $\Lambda$CDM simulations, and hence becomes the first known exception to the too--big--to--fail problem. This proves that these `massive failure' halos do exist in nature, confirming one of the key predictions of pure dark matter structure formation simulations in $\Lambda$CDM.

\item We use Jeans modelling to estimate the dark matter density profile of Tucana. Although the uncertainties are large, our results favour a high central density ($\rho_{\rm{DM}}$(150 pc)=$5.5\pm3.2\times10^8$ M$_{\odot}$ kpc$^{-3}$) and a halo mass $M_{200}=1.37^{+0.49}_{-0.44} \times 10^{10}$ M$_{\odot}$ consistent with abundance matching. Tucana's old--age population distinguishes it from other isolated, gas--rich galaxies which are still forming stars today. In models where dark matter is `heated' by baryonic feedback, Tucana is therefore expected to retain a higher central density than other isolated dwarfs. As anticipated from the lack of recent star formation, Tucana is consistent with residing in a `pristine' dark matter halo, unaffected by dark matter heating (see Fig. \ref{fig:dm_heating}). Further spectroscopic follow--up, particularly in the poorly sampled central regions of the galaxy, would be required to confirm the presence of a potential cusp in the density profile.

\end{itemize}

\section*{Acknowledgements}
The authors would like to thank John Pritchard of ESO Operations Support for his assistance with the wavelength calibration of the spectra. Based on observations collected at the European Organisation for Astronomical Research in the Southern Hemisphere under ESO programme 095.B-0133(A). This paper includes data gathered with the 6.5m Magellan Telescopes located at Las Campanas Observatory, Chile. This work makes use of PyAstronomy and \texttt{emcee}. DRW is supported by a fellowship from the Alfred P. Sloan Foundation. NFM acknowledges that this work has been published under the framework of the IdEx Unistra and benefits from a funding from the state managed by the French National Research Agency as part of the investments for the future program. NFM also acknowledge support by the Programme National Cosmology et Galaxies (PNCG) of CNRS/INSU with INP and IN2P3, co--funded by CEA and CNES.




\bibliographystyle{mnras}
\bibliography{tucana} 

\begin{thebibliography}{}
\makeatletter
\relax
\def\mn@urlcharsother{\let\do\@makeother \do\$\do\&\do\#\do\^\do\_\do\%\do\~}
\def\mn@doi{\begingroup\mn@urlcharsother \@ifnextchar [ {\mn@doi@}
  {\mn@doi@[]}}
\def\mn@doi@[#1]#2{\def\@tempa{#1}\ifx\@tempa\@empty \href
  {http://dx.doi.org/#2} {doi:#2}\else \href {http://dx.doi.org/#2} {#1}\fi
  \endgroup}
\def\mn@eprint#1#2{\mn@eprint@#1:#2::\@nil}
\def\mn@eprint@arXiv#1{\href {http://arxiv.org/abs/#1} {{\tt arXiv:#1}}}
\def\mn@eprint@dblp#1{\href {http://dblp.uni-trier.de/rec/bibtex/#1.xml}
  {dblp:#1}}
\def\mn@eprint@#1:#2:#3:#4\@nil{\def\@tempa {#1}\def\@tempb {#2}\def\@tempc
  {#3}\ifx \@tempc \@empty \let \@tempc \@tempb \let \@tempb \@tempa \fi \ifx
  \@tempb \@empty \def\@tempb {arXiv}\fi \@ifundefined
  {mn@eprint@\@tempb}{\@tempb:\@tempc}{\expandafter \expandafter \csname
  mn@eprint@\@tempb\endcsname \expandafter{\@tempc}}}

\bibitem[\protect\citeauthoryear{{Aparicio} et~al.,}{{Aparicio}
  et~al.}{2016}]{aparicio16}
{Aparicio} A.,  et~al., 2016, \mn@doi [\apj] {10.3847/0004-637X/823/1/9}, \href
  {https://ui.adsabs.harvard.edu/#abs/2016ApJ...823....9A} {823, 9}

\bibitem[\protect\citeauthoryear{{Avila-Vergara}, {Carigi}, {Hidalgo}  \&
  {Durazo}}{{Avila-Vergara} et~al.}{2016}]{avilavergara16}
{Avila-Vergara} N.,  {Carigi} L.,  {Hidalgo} S.~L.,   {Durazo} R.,  2016,
  \mn@doi [\mnras] {10.1093/mnras/stw205}, \href
  {http://adsabs.harvard.edu/abs/2016MNRAS.457.4012A} {457, 4012}

\bibitem[\protect\citeauthoryear{{Bermejo-Climent} et~al.,}{{Bermejo-Climent}
  et~al.}{2018}]{bermejo18}
{Bermejo-Climent} J.~R.,  et~al., 2018, preprint, \href
  {http://adsabs.harvard.edu/abs/2018arXiv180607679B} {} (\mn@eprint {arXiv}
  {1806.07679})

\bibitem[\protect\citeauthoryear{{Bernard} et~al.,}{{Bernard}
  et~al.}{2009}]{bernard09}
{Bernard} E.~J.,  et~al., 2009, \mn@doi [\apj] {10.1088/0004-637X/699/2/1742},
  \href {http://adsabs.harvard.edu/abs/2009ApJ...699.1742B} {699, 1742}

\bibitem[\protect\citeauthoryear{{Binney} \& {Mamon}}{{Binney} \&
  {Mamon}}{1982}]{binney82}
{Binney} J.,  {Mamon} G.~A.,  1982, \mn@doi [\mnras] {10.1093/mnras/200.2.361},
  \href {http://adsabs.harvard.edu/abs/1982MNRAS.200..361B} {200, 361}

\bibitem[\protect\citeauthoryear{{Boylan-Kolchin}, {Bullock}  \&
  {Kaplinghat}}{{Boylan-Kolchin} et~al.}{2011}]{boylankolchin11}
{Boylan-Kolchin} M.,  {Bullock} J.~S.,   {Kaplinghat} M.,  2011, \mn@doi
  [\mnras] {10.1111/j.1745-3933.2011.01074.x}, \href
  {http://adsabs.harvard.edu/abs/2011MNRAS.415L..40B} {415, L40}

\bibitem[\protect\citeauthoryear{{Boylan-Kolchin}, {Bullock}  \&
  {Kaplinghat}}{{Boylan-Kolchin} et~al.}{2012}]{boylankolchin12}
{Boylan-Kolchin} M.,  {Bullock} J.~S.,   {Kaplinghat} M.,  2012, \mn@doi
  [\mnras] {10.1111/j.1365-2966.2012.20695.x}, \href
  {http://adsabs.harvard.edu/abs/2012MNRAS.422.1203B} {422, 1203}

\bibitem[\protect\citeauthoryear{{Brook} \& {Di Cintio}}{{Brook} \& {Di
  Cintio}}{2015}]{brook15}
{Brook} C.~B.,  {Di Cintio} A.,  2015, \mn@doi [\mnras] {10.1093/mnras/stv864},
  \href {http://adsabs.harvard.edu/abs/2015MNRAS.450.3920B} {450, 3920}

\bibitem[\protect\citeauthoryear{{Bullock} \& {Boylan-Kolchin}}{{Bullock} \&
  {Boylan-Kolchin}}{2017}]{bullock17}
{Bullock} J.~S.,  {Boylan-Kolchin} M.,  2017, \mn@doi [Annual Review of
  Astronomy and Astrophysics] {10.1146/annurev-astro-091916-055313}, \href
  {https://ui.adsabs.harvard.edu/#abs/2017ARA&A..55..343B} {55, 343}

\bibitem[\protect\citeauthoryear{{Castellani}, {Marconi}  \&
  {Buonanno}}{{Castellani} et~al.}{1996}]{castellani96}
{Castellani} M.,  {Marconi} G.,   {Buonanno} R.,  1996, \aap, \href
  {http://adsabs.harvard.edu/abs/1996A%26A...310..715C} {310, 715}

\bibitem[\protect\citeauthoryear{{Collins} et~al.,}{{Collins}
  et~al.}{2013}]{collins13}
{Collins} M.~L.~M.,  et~al., 2013, \mn@doi [\apj]
  {10.1088/0004-637X/768/2/172}, \href
  {http://adsabs.harvard.edu/abs/2013ApJ...768..172C} {768, 172}

\bibitem[\protect\citeauthoryear{{Collins} et~al.,}{{Collins}
  et~al.}{2014}]{collins14}
{Collins} M.~L.~M.,  et~al., 2014, \mn@doi [\apj] {10.1088/0004-637X/783/1/7},
  \href {http://adsabs.harvard.edu/abs/2014ApJ...783....7C} {783, 7}

\bibitem[\protect\citeauthoryear{{Di Cintio}, {Brook}, {Macci{\`o}}, {Stinson},
  {Knebe}, {Dutton}  \& {Wadsley}}{{Di Cintio} et~al.}{2014}]{dicintio14}
{Di Cintio} A.,  {Brook} C.~B.,  {Macci{\`o}} A.~V.,  {Stinson} G.~S.,  {Knebe}
  A.,  {Dutton} A.~A.,   {Wadsley} J.,  2014, \mn@doi [\mnras]
  {10.1093/mnras/stt1891}, \href
  {https://ui.adsabs.harvard.edu/#abs/2014MNRAS.437..415D} {437, 415}

\bibitem[\protect\citeauthoryear{{Dotter}, {Chaboyer}, {Jevremovi{\'c}},
  {Kostov}, {Baron}  \& {Ferguson}}{{Dotter} et~al.}{2008}]{dotter08}
{Dotter} A.,  {Chaboyer} B.,  {Jevremovi{\'c}} D.,  {Kostov} V.,  {Baron} E.,
  {Ferguson} J.~W.,  2008, \mn@doi [\apjs] {10.1086/589654}, \href
  {http://adsabs.harvard.edu/abs/2008ApJS..178...89D} {178, 89}

\bibitem[\protect\citeauthoryear{{Dubinski} \& {Carlberg}}{{Dubinski} \&
  {Carlberg}}{1991}]{dubinski91}
{Dubinski} J.,  {Carlberg} R.~G.,  1991, \mn@doi [\apj] {10.1086/170451}, \href
  {https://ui.adsabs.harvard.edu/#abs/1991ApJ...378..496D} {378, 496}

\bibitem[\protect\citeauthoryear{{Dutton} \& {Macci{\`o}}}{{Dutton} \&
  {Macci{\`o}}}{2014}]{dutton14}
{Dutton} A.~A.,  {Macci{\`o}} A.~V.,  2014, \mn@doi [\mnras]
  {10.1093/mnras/stu742}, \href
  {https://ui.adsabs.harvard.edu/#abs/2014MNRAS.441.3359D} {441, 3359}

\bibitem[\protect\citeauthoryear{{Errani}, {Pe{\~n}arrubia}  \&
  {Walker}}{{Errani} et~al.}{2018}]{errani18}
{Errani} R.,  {Pe{\~n}arrubia} J.,   {Walker} M.~G.,  2018, \mn@doi [\mnras]
  {10.1093/mnras/sty2505}, \href
  {https://ui.adsabs.harvard.edu/\#abs/2018MNRAS.481.5073E} {481, 5073}

\bibitem[\protect\citeauthoryear{{Foreman-Mackey}, {Hogg}, {Lang}  \&
  {Goodman}}{{Foreman-Mackey} et~al.}{2013}]{foremanmackey13}
{Foreman-Mackey} D.,  {Hogg} D.~W.,  {Lang} D.,   {Goodman} J.,  2013, \mn@doi
  [\pasp] {10.1086/670067}, \href
  {http://adsabs.harvard.edu/abs/2013PASP..125..306F} {125, 306}

\bibitem[\protect\citeauthoryear{{Fraternali}, {Tolstoy}, {Irwin}  \&
  {Cole}}{{Fraternali} et~al.}{2009}]{fraternali09}
{Fraternali} F.,  {Tolstoy} E.,  {Irwin} M.~J.,   {Cole} A.~A.,  2009, \mn@doi
  [\aap] {10.1051/0004-6361/200810830}, \href
  {http://adsabs.harvard.edu/abs/2009A%26A...499..121F} {499, 121}

\bibitem[\protect\citeauthoryear{{Gallart} et~al.,}{{Gallart}
  et~al.}{2015}]{gallart15}
{Gallart} C.,  et~al., 2015, \mn@doi [\apjl] {10.1088/2041-8205/811/2/L18},
  \href {http://adsabs.harvard.edu/abs/2015ApJ...811L..18G} {811, L18}

\bibitem[\protect\citeauthoryear{{Gatto}, {Fraternali}, {Read}, {Marinacci},
  {Lux}  \& {Walch}}{{Gatto} et~al.}{2013}]{gatto13}
{Gatto} A.,  {Fraternali} F.,  {Read} J.~I.,  {Marinacci} F.,  {Lux} H.,
  {Walch} S.,  2013, \mn@doi [\mnras] {10.1093/mnras/stt896}, \href
  {https://ui.adsabs.harvard.edu/\#abs/2013MNRAS.433.2749G} {433, 2749}

\bibitem[\protect\citeauthoryear{{Hidalgo} et~al.,}{{Hidalgo}
  et~al.}{2013}]{hidalgo13}
{Hidalgo} S.~L.,  et~al., 2013, \mn@doi [\apj] {10.1088/0004-637X/778/2/103},
  \href {https://ui.adsabs.harvard.edu/#abs/2013ApJ...778..103H} {778, 103}

\bibitem[\protect\citeauthoryear{{Higgs} et~al.,}{{Higgs}
  et~al.}{2016}]{higgs16}
{Higgs} C.~R.,  et~al., 2016, \mn@doi [\mnras] {10.1093/mnras/stw257}, \href
  {https://ui.adsabs.harvard.edu/#abs/2016MNRAS.458.1678H} {458, 1678}

\bibitem[\protect\citeauthoryear{{Jeans}}{{Jeans}}{1922}]{jeans1922}
{Jeans} J.~H.,  1922, \mn@doi [\mnras] {10.1093/mnras/82.3.122}, \href
  {http://adsabs.harvard.edu/abs/1922MNRAS..82..122J} {82, 122}

\bibitem[\protect\citeauthoryear{{Jiang} \& {van den Bosch}}{{Jiang} \& {van
  den Bosch}}{2015}]{jiang15}
{Jiang} F.,  {van den Bosch} F.~C.,  2015, \mn@doi [\mnras]
  {10.1093/mnras/stv1871}, \href
  {http://adsabs.harvard.edu/abs/2015MNRAS.453.3575J} {453, 3575}

\bibitem[\protect\citeauthoryear{{Kacharov} et~al.,}{{Kacharov}
  et~al.}{2017}]{kacharov17}
{Kacharov} N.,  et~al., 2017, \mn@doi [\mnras] {10.1093/mnras/stw3188}, \href
  {http://adsabs.harvard.edu/abs/2017MNRAS.466.2006K} {466, 2006}

\bibitem[\protect\citeauthoryear{{Kirby}, {Bullock}, {Boylan-Kolchin},
  {Kaplinghat}  \& {Cohen}}{{Kirby} et~al.}{2014}]{kirby14}
{Kirby} E.~N.,  {Bullock} J.~S.,  {Boylan-Kolchin} M.,  {Kaplinghat} M.,
  {Cohen} J.~G.,  2014, \mn@doi [\mnras] {10.1093/mnras/stu025}, \href
  {http://adsabs.harvard.edu/abs/2014MNRAS.439.1015K} {439, 1015}

\bibitem[\protect\citeauthoryear{{Koposov} et~al.,}{{Koposov}
  et~al.}{2011}]{koposov11}
{Koposov} S.~E.,  et~al., 2011, \mn@doi [\apj] {10.1088/0004-637X/736/2/146},
  \href {http://adsabs.harvard.edu/abs/2011ApJ...736..146K} {736, 146}

\bibitem[\protect\citeauthoryear{{Lavery} \& {Mighell}}{{Lavery} \&
  {Mighell}}{1992}]{lavery92}
{Lavery} R.~J.,  {Mighell} K.~J.,  1992, \mn@doi [\aj] {10.1086/116042}, \href
  {http://adsabs.harvard.edu/abs/1992AJ....103...81L} {103, 81}

\bibitem[\protect\citeauthoryear{{Lewis}, {Ibata}, {Chapman}, {McConnachie},
  {Irwin}, {Tolstoy}  \& {Tanvir}}{{Lewis} et~al.}{2007}]{lewis07}
{Lewis} G.~F.,  {Ibata} R.~A.,  {Chapman} S.~C.,  {McConnachie} A.,  {Irwin}
  M.~J.,  {Tolstoy} E.,   {Tanvir} N.~R.,  2007, \mn@doi [\mnras]
  {10.1111/j.1365-2966.2007.11395.x}, \href
  {http://adsabs.harvard.edu/abs/2007MNRAS.375.1364L} {375, 1364}

\bibitem[\protect\citeauthoryear{{Mamon} \& {{\L}okas}}{{Mamon} \&
  {{\L}okas}}{2005}]{mamon05}
{Mamon} G.~A.,  {{\L}okas} E.~L.,  2005, \mn@doi [\mnras]
  {10.1111/j.1365-2966.2005.09225.x}, \href
  {http://adsabs.harvard.edu/abs/2005MNRAS.362...95M} {362, 95}

\bibitem[\protect\citeauthoryear{{Martin} \& {Jin}}{{Martin} \&
  {Jin}}{2010}]{martin10}
{Martin} N.~F.,  {Jin} S.,  2010, \mn@doi [\apj]
  {10.1088/0004-637X/721/2/1333}, \href
  {http://adsabs.harvard.edu/abs/2010ApJ...721.1333M} {721, 1333}

\bibitem[\protect\citeauthoryear{{Martin}, {Ibata}, {Chapman}, {Irwin}  \&
  {Lewis}}{{Martin} et~al.}{2007}]{martin07}
{Martin} N.~F.,  {Ibata} R.~A.,  {Chapman} S.~C.,  {Irwin} M.,   {Lewis} G.~F.,
   2007, \mn@doi [\mnras] {10.1111/j.1365-2966.2007.12055.x}, \href
  {http://adsabs.harvard.edu/abs/2007MNRAS.380..281M} {380, 281}

\bibitem[\protect\citeauthoryear{{Mateo}}{{Mateo}}{1998}]{mateo98}
{Mateo} M.~L.,  1998, \mn@doi [Annual Review of Astronomy and Astrophysics]
  {10.1146/annurev.astro.36.1.435}, \href
  {https://ui.adsabs.harvard.edu/#abs/1998ARA&A..36..435M} {36, 435}

\bibitem[\protect\citeauthoryear{{McConnachie}}{{McConnachie}}{2012}]{mcconnachie12}
{McConnachie} A.~W.,  2012, \mn@doi [\aj] {10.1088/0004-6256/144/1/4}, \href
  {http://adsabs.harvard.edu/abs/2012AJ....144....4M} {144, 4}

\bibitem[\protect\citeauthoryear{{Merrifield} \& {Kent}}{{Merrifield} \&
  {Kent}}{1990}]{merrifield90}
{Merrifield} M.~R.,  {Kent} S.~M.,  1990, \mn@doi [\aj] {10.1086/115438}, \href
  {http://adsabs.harvard.edu/abs/1990AJ.....99.1548M} {99, 1548}

\bibitem[\protect\citeauthoryear{{Monelli} et~al.,}{{Monelli}
  et~al.}{2010}]{monelli10}
{Monelli} M.,  et~al., 2010, \mn@doi [\apj] {10.1088/0004-637X/722/2/1864},
  \href {http://adsabs.harvard.edu/abs/2010ApJ...722.1864M} {722, 1864}

\bibitem[\protect\citeauthoryear{{Moore}}{{Moore}}{1994}]{moore94}
{Moore} B.,  1994, \mn@doi [\nat] {10.1038/370629a0}, \href
  {http://adsabs.harvard.edu/abs/1994Natur.370..629M} {370, 629}

\bibitem[\protect\citeauthoryear{{Navarro}, {Frenk}  \& {White}}{{Navarro}
  et~al.}{1996}]{navarro96}
{Navarro} J.~F.,  {Frenk} C.~S.,   {White} S.~D.~M.,  1996, \mn@doi [\apj]
  {10.1086/177173}, \href {http://adsabs.harvard.edu/abs/1996ApJ...462..563N}
  {462, 563}

\bibitem[\protect\citeauthoryear{{O{\~n}orbe}, {Boylan-Kolchin}, {Bullock},
  {Hopkins}, {Kere{\v{s}}}, {Faucher-Gigu{\`e}re}, {Quataert}  \&
  {Murray}}{{O{\~n}orbe} et~al.}{2015}]{onorbe15}
{O{\~n}orbe} J.,  {Boylan-Kolchin} M.,  {Bullock} J.~S.,  {Hopkins} P.~F.,
  {Kere{\v{s}}} D.,  {Faucher-Gigu{\`e}re} C.-A.,  {Quataert} E.,   {Murray}
  N.,  2015, \mn@doi [\mnras] {10.1093/mnras/stv2072}, \href
  {https://ui.adsabs.harvard.edu/#abs/2015MNRAS.454.2092O} {454, 2092}

\bibitem[\protect\citeauthoryear{{Oosterloo}, {Da Costa}  \&
  {Staveley-Smith}}{{Oosterloo} et~al.}{1996}]{oosterloo96}
{Oosterloo} T.,  {Da Costa} G.~S.,   {Staveley-Smith} L.,  1996, \mn@doi [\aj]
  {10.1086/118155}, \href {http://adsabs.harvard.edu/abs/1996AJ....112.1969O}
  {112, 1969}

\bibitem[\protect\citeauthoryear{{Plummer}}{{Plummer}}{1911}]{plummer11}
{Plummer} H.~C.,  1911, \mn@doi [\mnras] {10.1093/mnras/71.5.460}, \href
  {http://adsabs.harvard.edu/abs/1911MNRAS..71..460P} {71, 460}

\bibitem[\protect\citeauthoryear{{Pontzen} \& {Governato}}{{Pontzen} \&
  {Governato}}{2012}]{pontzen12}
{Pontzen} A.,  {Governato} F.,  2012, \mn@doi [\mnras]
  {10.1111/j.1365-2966.2012.20571.x}, \href
  {https://ui.adsabs.harvard.edu/#abs/2012MNRAS.421.3464P} {421, 3464}

\bibitem[\protect\citeauthoryear{{Pontzen} \& {Governato}}{{Pontzen} \&
  {Governato}}{2014}]{pontzen14}
{Pontzen} A.,  {Governato} F.,  2014, \mn@doi [\nat] {10.1038/nature12953},
  \href {https://ui.adsabs.harvard.edu/#abs/2014Natur.506..171P} {506, 171}

\bibitem[\protect\citeauthoryear{{Read} \& {Erkal}}{{Read} \&
  {Erkal}}{2018}]{read18c}
{Read} J.~I.,  {Erkal} D.,  2018, arXiv e-prints, \href
  {https://ui.adsabs.harvard.edu/\#abs/2018arXiv180707093R} {p.
  arXiv:1807.07093}

\bibitem[\protect\citeauthoryear{{Read} \& {Gilmore}}{{Read} \&
  {Gilmore}}{2005}]{read05}
{Read} J.~I.,  {Gilmore} G.,  2005, \mn@doi [\mnras]
  {10.1111/j.1365-2966.2004.08424.x}, \href
  {https://ui.adsabs.harvard.edu/#abs/2005MNRAS.356..107R} {356, 107}

\bibitem[\protect\citeauthoryear{{Read} \& {Steger}}{{Read} \&
  {Steger}}{2017}]{read17b}
{Read} J.~I.,  {Steger} P.,  2017, \mn@doi [\mnras] {10.1093/mnras/stx1798},
  \href {http://adsabs.harvard.edu/abs/2017MNRAS.471.4541R} {471, 4541}

\bibitem[\protect\citeauthoryear{{Read}, {Wilkinson}, {Evans}, {Gilmore}  \&
  {Kleyna}}{{Read} et~al.}{2006}]{read06b}
{Read} J.~I.,  {Wilkinson} M.~I.,  {Evans} N.~W.,  {Gilmore} G.,   {Kleyna}
  J.~T.,  2006, \mn@doi [\mnras] {10.1111/j.1365-2966.2005.09959.x}, \href
  {http://adsabs.harvard.edu/abs/2006MNRAS.367..387R} {367, 387}

\bibitem[\protect\citeauthoryear{{Read}, {Agertz}  \& {Collins}}{{Read}
  et~al.}{2016}]{read16}
{Read} J.~I.,  {Agertz} O.,   {Collins} M.~L.~M.,  2016, \mn@doi [\mnras]
  {10.1093/mnras/stw713}, \href
  {http://adsabs.harvard.edu/abs/2016MNRAS.459.2573R} {459, 2573}

\bibitem[\protect\citeauthoryear{{Read}, {Iorio}, {Agertz}  \&
  {Fraternali}}{{Read} et~al.}{2017}]{read17a}
{Read} J.~I.,  {Iorio} G.,  {Agertz} O.,   {Fraternali} F.,  2017, \mn@doi
  [\mnras] {10.1093/mnras/stx147}, \href
  {http://adsabs.harvard.edu/abs/2017MNRAS.467.2019R} {467, 2019}

\bibitem[\protect\citeauthoryear{{Read}, {Walker}  \& {Steger}}{{Read}
  et~al.}{2018}]{read18a}
{Read} J.~I.,  {Walker} M.~G.,   {Steger} P.,  2018, \mn@doi [\mnras]
  {10.1093/mnras/sty2286}, \href
  {https://ui.adsabs.harvard.edu/\#abs/2018MNRAS.481..860R} {481, 860}

\bibitem[\protect\citeauthoryear{{Read}, {Walker}  \& {Steger}}{{Read}
  et~al.}{2019}]{read18b}
{Read} J.~I.,  {Walker} M.~G.,   {Steger} P.,  2019, \mn@doi [\mnras]
  {10.1093/mnras/sty3404}, \href
  {https://ui.adsabs.harvard.edu/\#abs/2019MNRAS.484.1401R} {484, 1401}

\bibitem[\protect\citeauthoryear{{Richardson} et~al.,}{{Richardson}
  et~al.}{2011}]{richardson11}
{Richardson} J.~C.,  et~al., 2011, \mn@doi [\apj] {10.1088/0004-637X/732/2/76},
  \href {https://ui.adsabs.harvard.edu/#abs/2011ApJ...732...76R} {732, 76}

\bibitem[\protect\citeauthoryear{{Robin}, {Reyl{\'e}}, {Derri{\`e}re}  \&
  {Picaud}}{{Robin} et~al.}{2003}]{robin03}
{Robin} A.~C.,  {Reyl{\'e}} C.,  {Derri{\`e}re} S.,   {Picaud} S.,  2003,
  \mn@doi [\aap] {10.1051/0004-6361:20031117}, \href
  {http://adsabs.harvard.edu/abs/2003A%26A...409..523R} {409, 523}

\bibitem[\protect\citeauthoryear{{Sales}, {Navarro}, {Abadi}  \&
  {Steinmetz}}{{Sales} et~al.}{2007}]{sales07}
{Sales} L.~V.,  {Navarro} J.~F.,  {Abadi} M.~G.,   {Steinmetz} M.,  2007,
  \mn@doi [\mnras] {10.1111/j.1365-2966.2007.12026.x}, \href
  {http://adsabs.harvard.edu/abs/2007MNRAS.379.1475S} {379, 1475}

\bibitem[\protect\citeauthoryear{{Saviane}, {Held}  \& {Piotto}}{{Saviane}
  et~al.}{1996}]{saviane96}
{Saviane} I.,  {Held} E.~V.,   {Piotto} G.,  1996, \aap, \href
  {http://adsabs.harvard.edu/abs/1996A%26A...315...40S} {315, 40}

\bibitem[\protect\citeauthoryear{{Simon} \& {Geha}}{{Simon} \&
  {Geha}}{2007}]{simon07}
{Simon} J.~D.,  {Geha} M.,  2007, \mn@doi [\apj] {10.1086/521816}, \href
  {http://adsabs.harvard.edu/abs/2007ApJ...670..313S} {670, 313}

\bibitem[\protect\citeauthoryear{{Springel} et~al.,}{{Springel}
  et~al.}{2008}]{springel08}
{Springel} V.,  et~al., 2008, \mn@doi [\mnras]
  {10.1111/j.1365-2966.2008.14066.x}, \href
  {http://adsabs.harvard.edu/abs/2008MNRAS.391.1685S} {391, 1685}

\bibitem[\protect\citeauthoryear{{Teyssier}, {Johnston}  \&
  {Kuhlen}}{{Teyssier} et~al.}{2012}]{teyssier12}
{Teyssier} M.,  {Johnston} K.~V.,   {Kuhlen} M.,  2012, \mn@doi [\mnras]
  {10.1111/j.1365-2966.2012.21793.x}, \href
  {http://adsabs.harvard.edu/abs/2012MNRAS.426.1808T} {426, 1808}

\bibitem[\protect\citeauthoryear{{Tollerud} et~al.,}{{Tollerud}
  et~al.}{2012}]{tollerud12}
{Tollerud} E.~J.,  et~al., 2012, \mn@doi [\apj] {10.1088/0004-637X/752/1/45},
  \href {http://adsabs.harvard.edu/abs/2012ApJ...752...45T} {752, 45}

\bibitem[\protect\citeauthoryear{{Tollerud}, {Boylan-Kolchin}  \&
  {Bullock}}{{Tollerud} et~al.}{2014}]{tollerud14}
{Tollerud} E.~J.,  {Boylan-Kolchin} M.,   {Bullock} J.~S.,  2014, \mn@doi
  [\mnras] {10.1093/mnras/stu474}, \href
  {http://adsabs.harvard.edu/abs/2014MNRAS.440.3511T} {440, 3511}

\bibitem[\protect\citeauthoryear{{Tolstoy} et~al.,}{{Tolstoy}
  et~al.}{2004}]{tolstoy04}
{Tolstoy} E.,  et~al., 2004, \mn@doi [\apjl] {10.1086/427388}, \href
  {http://adsabs.harvard.edu/abs/2004ApJ...617L.119T} {617, L119}

\bibitem[\protect\citeauthoryear{{Vasiliev}}{{Vasiliev}}{2019}]{vasiliev19}
{Vasiliev} E.,  2019, \mn@doi [\mnras] {10.1093/mnras/sty2672}, \href
  {https://ui.adsabs.harvard.edu/\#abs/2019MNRAS.482.1525V} {482, 1525}

\bibitem[\protect\citeauthoryear{{Walker}, {Belokurov}, {Evans}, {Irwin},
  {Mateo}, {Olszewski}  \& {Gilmore}}{{Walker} et~al.}{2009a}]{walker09b}
{Walker} M.~G.,  {Belokurov} V.,  {Evans} N.~W.,  {Irwin} M.~J.,  {Mateo} M.,
  {Olszewski} E.~W.,   {Gilmore} G.,  2009a, \mn@doi [\apj]
  {10.1088/0004-637X/694/2/L144}, \href
  {https://ui.adsabs.harvard.edu/#abs/2009ApJ...694L.144W} {694, L144}

\bibitem[\protect\citeauthoryear{{Walker}, {Mateo}, {Olszewski},
  {Pe{\~n}arrubia}, {Wyn Evans}  \& {Gilmore}}{{Walker}
  et~al.}{2009b}]{walker09}
{Walker} M.~G.,  {Mateo} M.,  {Olszewski} E.~W.,  {Pe{\~n}arrubia} J.,  {Wyn
  Evans} N.,   {Gilmore} G.,  2009b, \mn@doi [\apj]
  {10.1088/0004-637X/704/2/1274}, \href
  {http://adsabs.harvard.edu/abs/2009ApJ...704.1274W} {704, 1274}

\bibitem[\protect\citeauthoryear{{Wang}, {Frenk}, {Navarro}, {Gao}  \&
  {Sawala}}{{Wang} et~al.}{2012}]{wang12}
{Wang} J.,  {Frenk} C.~S.,  {Navarro} J.~F.,  {Gao} L.,   {Sawala} T.,  2012,
  \mn@doi [\mnras] {10.1111/j.1365-2966.2012.21357.x}, \href
  {http://adsabs.harvard.edu/abs/2012MNRAS.424.2715W} {424, 2715}

\bibitem[\protect\citeauthoryear{{Whiting}, {Hau}  \& {Irwin}}{{Whiting}
  et~al.}{1999}]{whiting99}
{Whiting} A.~B.,  {Hau} G.~K.~T.,   {Irwin} M.,  1999, \mn@doi [\aj]
  {10.1086/301142}, \href {http://adsabs.harvard.edu/abs/1999AJ....118.2767W}
  {118, 2767}

\bibitem[\protect\citeauthoryear{{van der Marel}}{{van der
  Marel}}{1994}]{vandermarel94}
{van der Marel} R.~P.,  1994, \mn@doi [\mnras] {10.1093/mnras/270.2.271}, \href
  {http://adsabs.harvard.edu/abs/1994MNRAS.270..271V} {270, 271}

\makeatother
\end{thebibliography}



\appendix

\section{Modelling Tucana in SIDM} \label{appendixa}
\begin{figure}
	\centering
	\includegraphics [width=\linewidth] {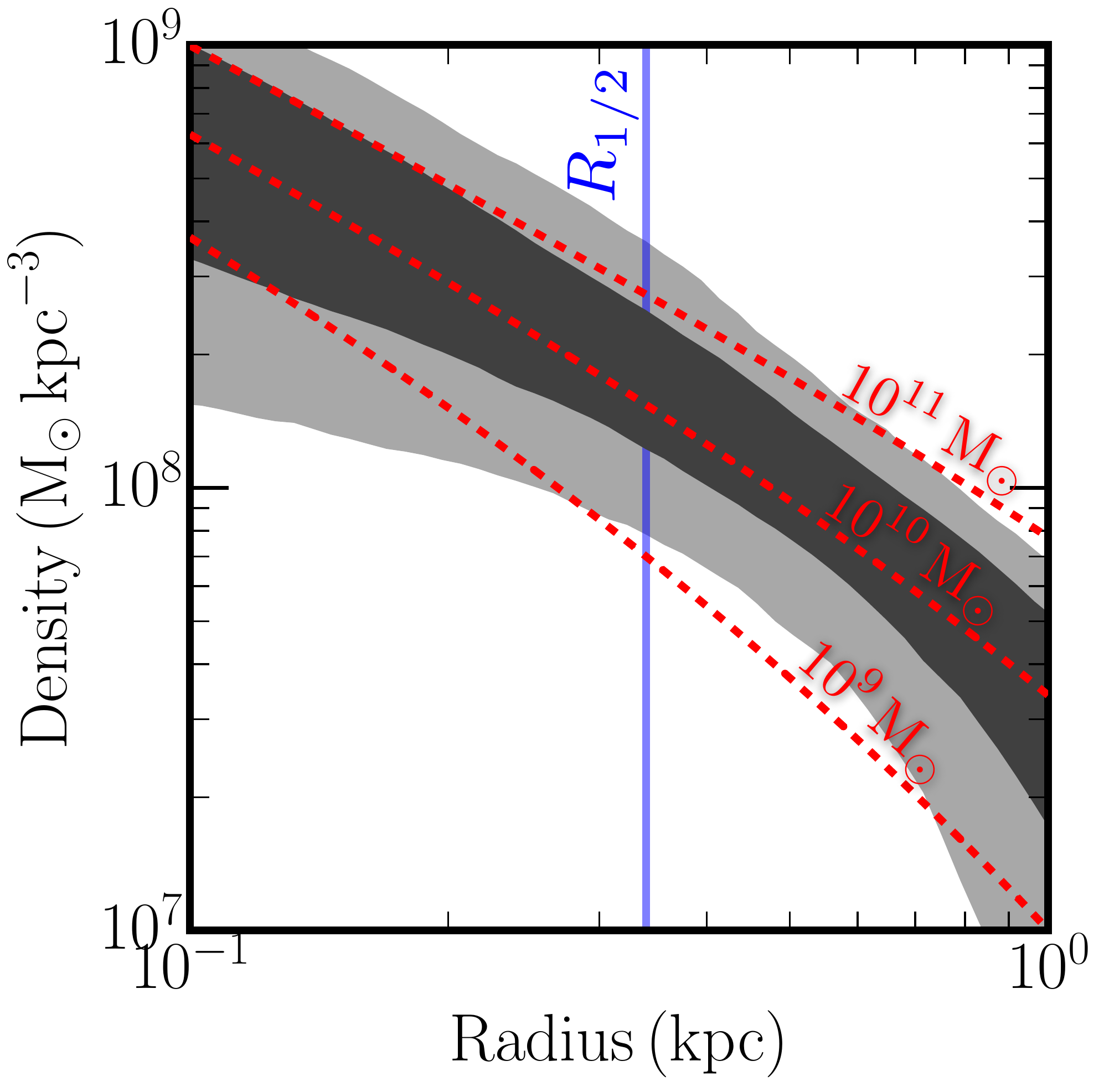}
	\caption{The density profile of Tucana, modelled using a \textsc{coreNFW} profile. Grey contours show the 1-- and 2--$\sigma$ uncertainty ranges. The results are in excellent agreement with our default `free-form' mass model (see Fig. \ref{fig:gravsphere_result}).}
	\label{fig:tucana_sidm}
\end{figure}

In this Appendix, we present the results of fitting Tucana with
a \textsc{coreNFW} profile (see \citealt{read16,read18a} for details of the functional form of this profile). This allows us to assess whether our results are sensitive to our choice of mass model for the dark matter halo of Tucana, and has the advantage that one of \textsc{coreNFW} parameters is the halo mass, M$_{200}$. The results, shown in Fig. \ref{fig:tucana_sidm}, are consistent with those of the free--form model (Fig. \ref{fig:gravsphere_result}). With \textsc{coreNFW}, we obtain a central density of $\rho_{\mathrm{DM}}$(150 pc)$=6.0_{-2.9}^{+3.7}\times 10^8$ M$_{\odot}$ kpc$^{-3}$ at the 68$\%$ confidence level. As was expected from \citet{read18a}, the density profile is systematically shallower in the innermost regions than with the free--form model. This effect is a result of the priors used, but is smaller than the measured uncertainties. The \textsc{coreNFW} model returns a pre--infall halo mass of $M_{200}=1.37^{+0.49}_{-0.44} \times 10^{10}$ M$_{\odot}$. Overall, the results of the two models are fully consistent with each other. This demonstrates that our inference of a high central density in Tucana is not dependent on our choice of mass model and priors.

\section{Mock Data Testing of \textsc{GravSphere}} \label{appendixb}

\begin{figure*}
	\centering
	\includegraphics[width=\linewidth]{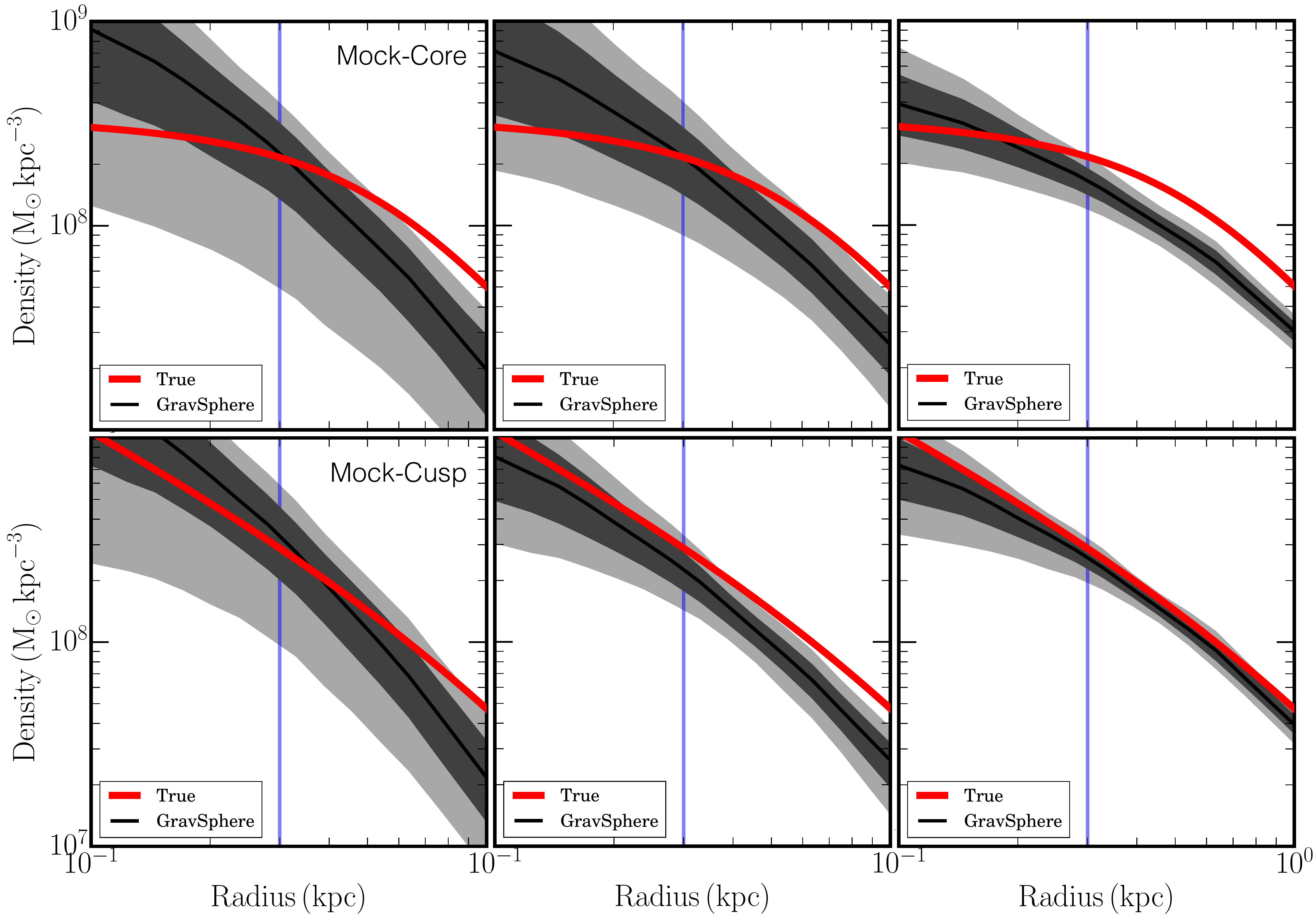}
	\caption{The results of our \textsc{GravSphere} mock data tests. The red line shows the underlying density profile, and the black line is the profile modelled by \textsc{GravSphere}, with contours showing 1-- and 2--$\sigma$ uncertainty ranges. Top row: Density profile for a cored mock galaxy, using 20 (left), 100 (middle), and 500 (right) probability--weighted effective mock members. Bottom row: Same, but for a cusped mock galaxy. The profile and central density are better recovered with an increased sample size. In the cored case, there is a slight bias towards lower densities at higher radii, though this does not affect the measurement of the central density.}
	\label{fig:gravsphere_mocks}
\end{figure*}

In this appendix we outline a series of mock data tests designed to investigate the effect of using \textsc{GravSphere} to model a small stellar sample.

We generate a series of mock data sets designed to reproduce the characteristics of the real kinematic data for Tucana. Each mock generates 2500 stars to reproduce the radial distribution of sources, the distribution of their velocity errors and their CMD positions. We set up the tracer density profile and kinematics for the mock assuming a Plummer light profile for the stars, with a stellar mass of $M_*=0.56\times10^6$ M$_\odot$ and a scale length of $r_P=0.284$ kpc, and a \textsc{coreNFW} profile for the dark matter halo with $M_{200}=10^10$ M$_\odot$ and a concentration parameter chosen to be twice the M$_{200}$--c$_{200}$ relation from \citet{dutton14} (to ensure a high velocity dispersion similar to that found in Tucana). The velocities for the stars were sampled from an isotropic distribution function generated using the AGAMA code \citep{vasiliev19}. Two mocks were constructed: one designed to represent a cusped galaxy; and one to represent a cored galaxy. The mocks include a foreground contribution mimicking any potential contamination of our real sample. Samples of $\sim$20, $\sim$100 and $\sim$500 effective members were then randomly drawn from the mock datasets. These samples represent the size of the true membership sample---with sampling designed to produce four membership--weighted radial bins as in the real Tucana model--- and two levels of increased sampling to illustrate the improvement achieved with more data.

\textsc{GravSphere} was then used to model the density profile of these mock data. The velocity dispersion profile $\sigma_{\rm{LOS}}$ was derived from the mock velocity data. To generate the surface brightness profile $\Sigma_{*}(R)$, mock photometry was produced, simulating the CMD positions and radial distribution of the Magellan/ MegaCam imaging catalogue. The resulting density profiles are shown for both the cusped and cored mocks in Fig. \ref{fig:gravsphere_mocks}. The red lines mark the underlying profile used to seed the mocks. In the cusped case (bottom row), the profile is well recovered by \textsc{GravSphere}, with the constraints tightening as the sample size increases. For the lowest-sampled mock (left panels), the recovered density profile has uncertainties similar to that for the real Tucana data (compare with Fig. \ref{fig:gravsphere_result}). In the cored case (top row), the central density is reasonably well recovered (within 1--$\sigma$), but the density profile is slightly biased towards lower densities than the input profile at large radii, such that the model appears cuspier than the true profile. The slight bias towards cusped profiles is noted and explored further in \citet{read18a}. There, it was shown that the bias diminishes with improved sampling to 1000 or 2000 effective members (their Figure B1). This is something that we will improve on in future work. The central density at 150pc-- which is of interest in this paper-- is well recovered, supporting our use of \textsc{GravSphere} to measure $\rho_{\rm{DM}}$(150pc) for Tucana.


\bsp	
\label{lastpage}
\end{document}